\documentclass[acmsmall, nonacm]{acmart}
\usepackage{multirow}
\usepackage{listings}
\usepackage{subcaption}
\usepackage[T1]{fontenc}
\usepackage{xspace}

\def\BibTeX{{\rm B\kern-.05em{\sc i\kern-.025em b}\kern-.08emT\kern-.1667em\lower.7ex\hbox{E}\kern-.125emX}}

\newcommand{\todo}[1]{{\textcolor{red}{[#1]}\normalfont}}

\newcommand{\system}{Co-ML\xspace}
\newcommand{\nonprofit}{KWK\xspace}
\newcommand{\nonprofitfull}{Kode with Klossy\xspace}
\newcommand{\campname}{AIML Summer Camp\xspace}

\begin{document}

\title[\system: Collaborative Machine Learning Model Building]{\system: Collaborative Machine Learning Model Building for Developing Dataset Design Practices} 

\author{Tiffany Tseng}
\email{tiffanytseng@apple.com}
\affiliation{\institution{Apple}}

\author{Matt J. Davidson}
\email{matt_davidson@apple.com}
\affiliation{\institution{Apple}}

\author{Luis Morales-Navarro*}
\thanks{* Work conducted at Apple}
\email{luismn@upenn.edu}
\affiliation{\institution{University of Pennsylvania}}

\author{Jennifer King Chen}
\email{jennkingchen@apple.com}
\affiliation{\institution{Apple}}

\author{Victoria Delaney*}
\email{vdoch@stanford.edu}
\affiliation{\institution{Stanford University}}

\author{Mark Leibowitz}
\email{mleibowitz@apple.com}
\affiliation{\institution{Apple}}

\author{Jazbo Beason}
\email{jontait_beason@apple.com}
\affiliation{\institution{Apple}}

\author{R. Benjamin Shapiro}
\email{nerd@apple.com}
\affiliation{\institution{Apple}}

\renewcommand{\shortauthors}{Tseng et al.}

\newcommand{\AbstractCategory}[1]{%
  \par\addvspace{.5\baselineskip}
  \noindent\textbf{#1}\quad\ignorespaces
}

%
\begin{abstract}
Machine learning (ML) models are fundamentally shaped by data, and building inclusive ML systems requires significant considerations around how to design representative datasets. Yet, few novice-oriented ML modeling tools are designed to foster hands-on learning of dataset design practices, including how to design for data diversity and inspect for data quality.

To this end, we outline a set of four data design practices (DDPs) for designing inclusive ML models and share how we designed a tablet-based application called \system to foster learning of DDPs through a collaborative ML model building experience. With \system, beginners can build image classifiers through a distributed experience where data is synchronized across multiple devices, enabling multiple users to iteratively refine ML datasets in discussion and coordination with their peers.

We deployed \system in a 2-week-long educational \campname, where youth ages 13-18 worked in groups to build custom ML-powered mobile applications. Our analysis reveals how multi-user model building with \system, in the context of student-driven projects created during the summer camp, supported development of DDPs including incorporating data diversity, evaluating model performance, and inspecting for data quality. Additionally, we found that students' attempts to improve model performance often prioritized learnability over class balance. Through this work, we highlight how the combination of collaboration, model testing interfaces, and student-driven projects can empower learners to actively engage in exploring the role of data in ML systems.
\end{abstract}

%

\ccsdesc[300]{Human-centered computing~Empirical studies in HCI}

%
\keywords{machine learning, collaboration, computing education, data science}
  
%
\maketitle

\section{Introduction}
Machine learning (ML) has dramatically impacted a wide-range of domains including healthcare, entertainment, and communication. As ML systems play an increasing role in our daily lives, there is a growing need for ML education to support people to engage purposefully with ML technologies. Improving ML literacy not only strengthens technical understanding of ML, but also promotes civic engagement, interest in ML, and skill-building for future creators of and with ML technologies. 

Several frameworks have been proposed for introducing ML in primary and secondary education, as well as for non-experts more broadly \cite{touretzky2022machine, long2020ai, zhou2020designing}. These frameworks propose foundational concepts people should learn about ML, including that data plays a central role in how ML models make decisions, and that humans play a role in shaping the ways data is selected. Yet these frameworks are initial guidelines, and research is needed to inform both teaching and tool design that can effectively support non-experts in learning these ideas. Even within higher education, prior work has stressed the importance of engaging with dataset considerations essential to model robustness and generalizability, which are often overlooked in favor of educating future ML practitioners about model architecture and the use of off-the-shelf datasets \cite{sambasivan2021everyone}. Because many harmful problems with ML applications today result from a lack of representative data \cite{buolamwini2018gender, eubanks2018automating}, educational efforts to increase understanding about the role of data in ML are especially needed. 

A growing number of tools have been created for beginners to build ML models using their own data  \cite{carney2020teachable, agassi2019scratch, tseng2021plushpal, zimmermann2019youth, tang2019empowering}. However, existing tools have limited support for learning to design datasets for inclusive models because they are,

\begin{enumerate}
    \item optimized for data collection and testing from a single user, increasing the likelihood of imbalanced or biased data that may not generalize to other use cases
    \item center on ephemeral live classification, where model results are evaluated in real time, lacking support for reviewing and debugging misclassifications over time and monitoring how model performance might change in response to refining data 
\end{enumerate}

To address these gaps, we developed a novel ML model-building tool called \system for multi-user data collection and model testing on tablets. We made collaboration a central feature of the \system experience because we envisioned that working with others to gather data, analyze data, and test models can surface multiple points of view beyond those an individual learner might consider on their own. Through a collective experience for reviewing, discussing, and debugging ML models and datasets, we expected that learners would consider, address, and enact key dataset design practices such as incorporating dataset diversity and inspecting for data quality as they built their models with \system.

In this paper, we describe how features of \system were designed with these dataset design practices in mind. We then share our evaluation of \system through a 2-week long ML Summer Camp pilot for high school girls and non-binary youth in partnership with the non-profit \nonprofitfull. In the camp, youth worked in groups to create ML models with \system and build their own mobile applications applying these models to personally relevant topics of their choice such as healthy eating and sustainable fashion. Throughout the camp, our research team studied how students worked together to build their ML applications. Our mixed methods analytical approach incorporates data sources such as observation notes and audio recordings of student interactions, in-app logs of how students navigated the app, design journals students updated while building their final projects, and post-camp semi-structured interviews to learn about challenges novices faced when designing and debugging ML models.

Through our analysis, we examine the following research question: How does collaboration supported by \system shape the learning of four dataset design practices: 1) incorporating dataset diversity, 2) evaluating model performance and its relationship to data, 3) balancing datasets, and 4) inspecting for data quality?

Our contributions are

\begin{enumerate}
    \item A framework of dataset design practices for learning about the role of data in ML systems
    \item A description of how features in \system, a novel, collaborative ML modeling app, along with an accompanying learning experience, were designed to foster the development of dataset design practices
    \item A discussion of how collaboration enabled by \system shaped student understanding of each dataset design practice through our deployment and study of \system in 2-week summer camps for teenagers
\end{enumerate}
\section{Related Work}

We begin with an overview of professional ML data work to provide context about data considerations in ML practice, followed by a summary of ML education efforts for youth and research on collaborative learning. Finally, we summarize the gaps in existing novice-oriented ML tools for fostering dataset design practices, which leads into our description of the \system application and how we designed a collaborative learning experience to address these gaps.

Our work focuses specifically on ML, a subset of AI in which computers detect patterns in data to make useful predictions about new data. In particular, we align with efforts to empower more people to contribute to ML (even without deep expertise in ML algorithms and architecture) by building supports for \textit{machine teaching} \cite{simard2017machine, ramos2020interactive}, which emphasizes human teachers and how they interact with data to build ML systems. In doing so, our goal is to enable beginners to reason about the role data plays in ML models by directly engaging in building datasets and models.

\subsection{The Role of Data in ML}
\label{sec:role-of-data-in-ml}

Data is foundational to building ML models and largely determines model performance, robustness, and generalizability \cite{sambasivan2021everyone}. Preparing data is an essential part of the ML modeling pipeline, estimated to take up 80\% of a data scientist's time \cite{redman2018if}. Data preparation involves many considerations across the model building process, from planning and prioritizing what data to initially collect, to iterating on data in response to model performance issues or shifting project goals \cite{hohman2020understanding}. Failing to properly design datasets fundamentally jeopardizes the usefulness of a model, as the classic adage "garbage in, garbage out" advises \cite{babbage2022passages}, and can lead to \textit{data cascades} or compounding negative downstream effects as a result of data issues \cite{sambasivan2021everyone}. Using unrepresentative data, or data that do not accurately reflect a model's intended users and use cases, can result in problematic, biased models that discriminate on a range of attributes including race and gender \cite{buolamwini2018gender, dastin2018amazon, noble2018algorithms}. 

In preparing data, model designers need to consider multiple dimensions of data quality that are context dependent \cite{tayi1998examining}. These dimensions include how \textit{representative} the data is, or how comprehensively the data reflects characteristics important for the model's ultimate use; how \textit{clean} the data is (that the data is properly labelled and does not contain duplicates, for example); how \textit{balanced} a dataset is, or that the distribution of samples across classes is equitable (the opposite of this is considered \textit{class imbalance}); and how \textit{diverse} the data is, or that the data reflects variability in the populations using a model and the contexts in which the model is used \cite{cdt, hopkins2023designing}. While ML practitioners can leverage different techniques to measure the diversity of their dataset \cite{mitchell2020diversity}, they also need to take into account \textit{learnability}, as some classes may be inherently harder for a model to learn and thus require more data than others \cite{hopkins2023designing}. 

In an effort to improve their models, practitioners often iterate on their data. Typically ML models include training, testing, and validation datasets, where training data is used to train a model, validation data is used to assess model performance when tuning model parameters, and testing data is for evaluating performance of a trained model. Iteration often includes adding, removing, or modifying samples from any of these datasets. Adding data can involve expanding the dataset by collecting additional samples from a random population, or more targeted efforts to address underrepresentation or class imbalance for a specific label; removing data happens in response to identifying noisy or erroneous data \cite{hohman2020understanding}. Prior work examining practitioners' sensemaking practices when reviewing data found that collaboration, where multiple people interact and discuss data, can help support decision making about and inform understanding of data \cite{koesten2021talking}.  

Despite the importance of data work in professional ML practice, ML education efforts, as well as ML research, have historically focused on model architecture rather than data design practices. Undergraduate and graduate courses in AI and ML often use existing toy datasets from Kaggle or similar platforms, limiting opportunities for students to actively learn about data preparation practices and data quality issues firsthand \cite{sambasivan2021everyone}. As we consider how AI and ML education might expand in service of even younger audiences in K-12, we see the need and opportunity for foundational data design practices to be incorporated into tools and learning activities so that learners can authentically encounter and engage with data as they learn how ML models work. 

\subsection{ML Education for Youth}
\label{sec:ml-education-k-12}

Efforts to build out educational resources for learning ML are motivated by the growing role ML technologies play in our daily lives, alongside issues like a lack of transparency and public understanding of how these systems work. While efforts to integrate AI education into school took place as early as the 1970s, there has been a marked growth in instructional resources over the past half decade \cite{marques2020teaching}, with a shift towards ML more specifically \cite{kahn2021constructionism}. The importance of ML education is further recognized by this journal specifically, with a special issue of TOCE from 2019 highlighting opportunities for creative applications in arts and design \cite{shapiro2019introduction, fiebrink2019machine}.

Over the last decade, a number of AI Literacy frameworks have been proposed, outlining educational opportunities for teaching how AI and ML systems work. These frameworks include the AI4K12 5 Big Ideas in Artificial Intelligence \cite{ai4k12}, AI Competencies \cite{long2020ai}, and design frameworks for K-12 AI Education \cite{zhou2020designing}. Within these frameworks, there are several core considerations around data in ML models, including that computers learn from data (Big Idea 3 from \cite{ai4k12} and Competency 12 from \cite{long2020ai}), and that training datasets are often constructed and edited by humans and affects how models perform (Competency 13 \cite{long2020ai}). These frameworks also stress the importance of accounting for bias that can result when populations are underrepresented in training data, and that humans play a role in dataset construction, cleaning, and verification that can minimize these biases. Arastoopour Irgens and colleagues, for example, \cite{irgens2022characterizing, irgens2022designing} investigated how children (aged 9-13) used Google Teachable Machine to build and critique models, guided by curriculum that engaged them in considering the impact of biased datasets as well as the social, ethical, and political implications of ML technologies. Because ML systems have fundamental differences compared to classical notional machines, Tedre et al. have argued that computational thinking frameworks need to be expanded to accommodate ML practices and ideas such as inductive problem solving techniques and trial and error approaches to debugging \cite{tedre2021ct}.

The AI4K12 guidelines around Learning (Big Ideas 3) underscore the need for \textit{big data}, describing how ``Large amounts of training data are required`` along with ``thousands to millions of trial and error experiments to solve narrowly defined problems.`` However, transfer learning is enabling people to build functioning models on smaller datasets by modifying pre-trained models \cite{weiss2016survey, smalldata2020}. Young learners can effectively use transfer learning to iteratively improve models by analyzing misclassifications and revising datasets quickly in the context of small projects \cite{hitron2018introducing, vartiainen2020machine}. Transfer learning presents opportunities for novices to use their own data to build and evaluate ML models rather than relying on the use of existing large datasets; prior work has also found that learners using personally-relevant data were able to reason about the mechanisms of an ML-system and self-advocate against potentially harmful results \cite{register2020learning}. 

While proposed AI Literacy frameworks put forth ideas for age-appropriate ML concepts, research is needed to validate the appropriateness of these frameworks and to determine practices, tools, and approaches that can best support learning of these ideas. A growing number curricular tools and resources have been developed for age-appropriate introductions to AI \cite{marques2020teaching, druga2022landscape}. Existing efforts aimed at youth at the secondary education level (aged 11-18) often forefront ethics \cite{payne2019ethics, williams2022ai, ali2019constructionism, lee2021developing, lee2022black}, centering on the use and exploration of existing ML models. In contrast to using off-the-shelf ML models, our work focuses on the design and implementation of ML tools that allow learners to create ML models and projects using data they collect themselves.

Several novice-oriented tools for ML model building extend popular blocks-based programming environments to accommodate ML features, including extensions to Scratch \cite{williams2022ai}, MIT App Inventor \cite{tang2019empowering, van2021teaching}, and Snap! \cite{kahn2021constructionism}. These examples typically combine off-the-shelf pre-trained models with interfaces for beginners to add their own data and train a model \cite{druga2018growing, lane2021machine}, supporting novices building applications with features like sentiment analysis and image detection. Other novice-oriented ML tools include production applications like Google Teachable Machine \cite{carney2020teachable} and prototype research tools like PlushPal \cite{tseng2021plushpal} and AlpacaML \cite{zimmermann2020youth, zimmermann2019youth}, where beginners can create image, gesture, or sound recognition models. Typically, novice-oriented ML modeling tools incorporate live classification interfaces for evaluating a model; for example, for an image classifier, the interface may show in real time what the model's top prediction is based on live streaming camera input. Evaluations of these tools suggest that when students use ML tools to create personally meaningful projects, they are able to demonstrate increased understandings of ML concepts \cite{ali2019constructionism, druga2018growing, estevez2019gentle, jordan2021poseblocks, tang2019empowering, dwivedi2021exploring, kahn2021constructionism, vartiainen2020machine, zimmermann2019youth}. 

Despite these affordances, we identified several gaps in existing ML modeling tools. First, these tools center around individual use, where a single person trains and tests a model using only their own data. Having data from a single user limits opportunities to work with diverse data representing use cases and perspectives beyond what an individual may consider on their own; important dataset considerations like dataset diversity and model generalization may thus be more difficult to encounter and resolve. 

Second, because existing tools center model evaluation on real-time live classification, they lack support for systematic review of misclassified examples and assessment of model improvement in response to changes (such as adding different training data or tuning model parameters over repeated iterations). Typically, in professional ML practice, models are assessed with regard to test datasets, a portion of collected data specially reserved for evaluation; yet, beginner-friendly ML tools today lack support for test datasets, making it more difficult for users to assess whether or not their model is improving. 

Finally, existing tools are largely web-based and are not supported on mobile devices. Mobile-supported data collection could more flexibly accommodate learners collecting data in the wild (as compared to front-facing laptop webcams for images and video, for example), which may further support opportunities for expanding dataset diversity.

\subsection{Collaborative Learning} 
Collaboration with peers has the potential to positively impact student learning in appropriately designed activities \cite{roschelle1992learning, blumenfeld1996learning}. Collaborative work can activate socio-cognitive processes for learning: asking questions, explaining one’s thinking, providing a critique, and resolving differing perspectives, all of which are difficult or impossible to accomplish when working alone \cite{barron2003smart}. The benefits of peer-to-peer collaboration are well-documented in computer science education \cite{roschelle1995construction, hmelo2017computer}, with prior work highlighting how, through collaboration, students can learn how to share responsibility while completing a task \cite{lytle2020investigating}, participate in and learn from group discussions \cite{porter2016multi}, and pair program with increased confidence \cite{bigman2021pearprogram, werner2004pair}. Students engaging in peer collaboration are also more likely to persist in the discipline \cite{braught2011case, salleh2011effects}. Further, prior work has found active discussion in peer-to-peer problem solving \cite{barron2003smart} to correlate with novelty of student-created designs \cite{deitrick2014discourse}.

Standard pair programming models typically involve students working alongside one another on a single machine, with one person "driving" the experience by having control over a code editor. Prior work on pair programming has explored how collaborative dialogue between students may positively shape how students debug and learn together \cite{exploringpairprogramming, jones2013use}, with a range of types of dialogue that may be more or less supportive \cite{deitrick2016we}.

In contrast to this notion of pair programming, our work focuses on \textit{synchronous co-editing}, where multiple people can edit a single project simultaneously on multiple devices. While synchronous editing has been examined in the context of collaborative blocks-based editors \cite{lytle2020investigating, selwyn2019co}, multi-user experiences in the context of data science and ML has focused almost exclusively on professionals. For example, recent work has contributed computational notebooks that support multi-user synchronous editing \cite{googleColab, wang2019data} for professional data scientists, finding that synchronous editors can increase group exploration and reduce communication costs compared to developers working in individual notebooks \cite{wang2019data}. More broadly, an important feature of collaborative software tools is improving group awareness so that team members can better coordinate actions and develop a shared mental model for their work \cite{dourish1992awareness}.

In the context of ML education, recent work has highlighted a need for research to examine the role collaboration can play in student learning \cite{sanusi2022exploring}. Studies conducted with families (adults and children) have shown affordances for group learning and reasoning about ML \cite{druga2022family, long2022family} by supporting parent-child dialogue and distinct parental facilitation roles. While some related work in ML education for youth described in Section \ref{sec:ml-education-k-12} involves students working in collaborative learning settings like such as classrooms \cite{vartiainen2020machine}, multi-day camps \cite{vachovsky2016toward}, workshops \cite{zimmermann2019youth}, and online interventions \cite{williams2022ai}, to the best of our knowledge, only two previous studies have specifically studied the nature and affordances of peer-to-peer student collaboration when working on ML projects. 

Prior work examining how youth use Google Teachable Machine described an activity where students built models individually using workshop-supplied objects, then swapped models with others for testing. This helped students realize situations where a model might not work well for new users \cite{dwivedi2021exploring}. Kaspersen and colleagues \cite{kaspersen2021votestratesml} developed VotestratesML, a web-based tool designed for students (aged 17-20) in social studies classes to explore machine learning in small groups. Using VostratratesML with pre-existing datasets, students collaboratively made decisions about tuning model parameters and making revisions based on choices of features, algorithms, and model output. Because groups’ results were projected publicly, students were more engaged in discussions about their results.

An area understudied in prior ML education research is how learners can collaboratively design models using their own constructed datasets, rather than supplied materials or existing toy datasets. Our work with \system explores the potential of this type of collaborative experience in the context of students building models addressing topics of personal relevance. 

\subsection{Research Opportunities for Fostering Dataset Design Practices for Novices} 

Our review of current ML education efforts identified that existing novice-oriented ML tools center around a single-user experience where an individual collects only their own data, providing limited opportunities for learners to create diverse, balanced datasets that meet the needs of more than a single individual or use case. Additionally, these tools lack robust model evaluation metrics and features that enable users to engage with key skills in professional ML practice, such as building test datasets and evaluating their model performance across multiple iterations of their datasets.

We argue that a collaborative modeling experience may help mitigate these issue because 1) larger datasets can be built more quickly by multiple users compared to a single user, potentially providing more opportunities for issues like imbalanced data to arise; 2) individual differences in data collection strategies may become visible through a multi-user experience, since people may have different points of view on what is considered to be representative data --- and this may, in turn, lead to more diverse data; and 3) a collaborative experience provides opportunities for collective discussion of model issues, which may help learners deepen their insights about how ML models work through active discussion with their peers. Our hope is that a collaborative ML modeling tool can help learners create diverse datasets and link the design of their datasets with the performance of a ML system.

Next, we summarize the specific dataset design practices we designed \system to support, followed by a description of the design of the \system experience.

\section{Dataset Design Practices}
\label{sec:dataset-design-practices}

Summarizing the considerations and practices of ML practitioners for collecting data (described in \ref{sec:role-of-data-in-ml} and by \cite{hohman2020understanding, sambasivan2021everyone, cdt, hopkins2023designing}), we propose a set of dataset design practices (DDPs) for fostering learning about the role of data in ML modeling. Specifically, this set of DDPs is intended to foster student learning as they collect and construct their own datasets in the context of building models for their own use. 

\begin{table}[h]
    \centering
    \begin{tabular}{ p{1cm} p{4cm} p{7cm} }
    \toprule
     & \textbf{Dataset Design Practice} & \textbf{Description} \\ \hline
    DDP1 & Incorporating dataset diversity & Ensuring that data is representative and accounts for the diverse characteristics of a label and the variety of use cases where a model might be used \\ \hline
    DDP2 & Evaluating model performance and its relationship to data & Understanding how well a model is performing
    \newline \newline Identifying gaps or confounding factors in data that might impact model performance 
    \newline \newline Assessing whether a model has improved after dataset revisions and model retraining \\ \hline
    DDP3 & Balancing datasets & Designing datasets that have roughly equal distribution of samples across labels and ensuring model performance is consistent across labels \\ \hline
    DDP4 & Inspecting for data quality & Checking that data is properly labelled and of sufficient quality (e.g., that image data is not blurry) \\ 
    \bottomrule
    \end{tabular}
    \caption{Dataset design practices}
    \label{tab:dataset-design-practices}
\end{table}

We will refer to Table \ref{tab:dataset-design-practices} as we describe how \system was designed to support these dataset design practices, particularly in the context of novices building supervised ML image classifiers. Further, we present the results of our evaluation of \system with respect to how students engaged with these DDPs as they designed custom ML models.

\section{\system System and Companion Starter App}
\label{sec:system}

To support beginners with thoughtful exploration of the role of data in ML model performance, we designed a collaborative modeling mobile app called \system. \system is a tablet-based app that supports a multi-user experience for collecting image data, training an ML image classifier, and testing the model’s performance. Multiple people can collect image data using the camera on their individual tablet, and the data is synchronized across devices so that everyone is working with the same shared dataset as they iterate on their models.

In this section, we share our design goals in creating \system along with a description of the app. Because our intention is to support users with integrating ML models into custom applications, we also describe a companion starter app we designed to be used alongside \system to create mobile apps integrating ML models. While analysis of students' use of the starter app is out of scope for our study, we share its design to provide context for interpreting our descriptions of what students built.

\subsection{\system Design Goals}

We had two primary design goals for the \system application to enable: 1) diverse perspectives in the model building process, and 2) iterative model testing for monitoring how model performance changes in response to dataset revisions. In this section, we describe how these two goals drive the design of the features of \system and how they relate to the dataset design practices (Section \ref{sec:dataset-design-practices}) we intend to foster.

For our first design goal of enabling diverse perspectives, we imagined that individuals may differ in the ways they collect data, including how they take photographs and the contexts in which they take them. We expected these differences to emerge when multiple people contribute to a dataset, especially with a mobile interface that allows flexible data collection in a variety of settings. As a result, a core consideration in the design of \system was how the data is synchronized and displayed throughout the modeling experience so that users are more likely to encounter other people’s perspectives and practices as they build their models. We designed \system to support learners in incorporating those differences by diversifying their dataset (DDP1), considering class balance as it relates to sample distribution (DDP3), and inspecting and assessing data quality from multiple individuals (DDP4).

For our second design goal of enabling iterative model testing, we designed \system to provide feedback about how well the model is working to best support learners exploring the relationship between data and model performance (DDP 2). To this end, the app provides actionable, beginner-friendly feedback to let users know how well their model works at the moment and whether the model has improved after retraining. We also recognize that iteration is easier when there is minimal friction in the model training process, so we aimed to reduce model training time and latency when synchronizing data across devices.  

\subsection{\system System Description}

\system has a data-focused ML model building experience that incorporates the following user flow: 1) defining an ontology of labels, 2) collecting data, 3) training a model, 4) evaluating model performance, and 5) iterating on a model by modifying the label ontology and or datasets. After creating a project, users can generate a shareable link (sent via email or text message, for example) and invite collaborators to contribute. Here, we walk through an example scenario involving multiple people collaborating on a project to identify various fruits, with each person using their own tablet to collect data.

\begin{figure}[h]
  \centering
  \includegraphics[width=\linewidth]{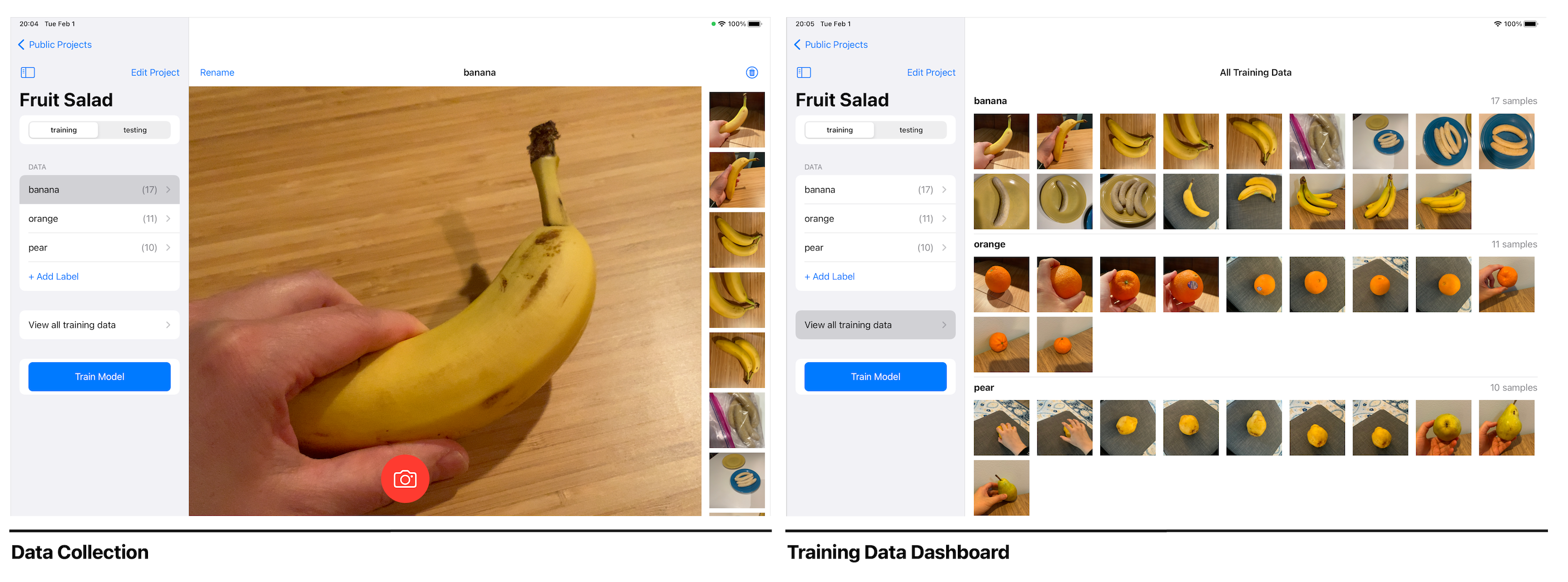}
  \caption{Data collection interface for adding labelled images to the shared dataset (left).  All images added across devices are visible in the synchronized Training Data Dashboard, organized by label name (right).}
  \Description{The Data collection interface and training data dashboard for Fruit Salad project}
  \label{fig:training-mode}
\end{figure}

\subsubsection{Define ontology of labels in a classifier}
Users provide the names of labels that their classifier will be able to predict. Any user can add, rename, or delete labels in a project.

\subsubsection{Collect Data}
Selecting a label from their list of labels opens up the tablet camera, where image data for that label can be captured.  When adding images, a stream of recently added images for a given label (consolidated from all users within the project) are displayed alongside the camera feed to bring awareness to what images have already been added and to encourage users to inspect different ways others might be capturing their training data. This is shown in Figure \ref{fig:training-mode}. Further, the number of images per label is displayed next to the label name so that users can monitor how their dataset is being updated and potentially consider labels that might need more attention due to a lack of data or imbalance (DDP3).

At any point, users can tap the View All Data button to view the Training Data Dashboard, which displays training data collected across all users, organized by label. The dashboard consists of a grid of training images, with 25 images visible at a time on a 11'' tablet screen to encourage users to look for patterns or gaps in their dataset and inspect for any issues around data quality (DDP4). To support users inspecting lots of data at once, we designed \system for tablets as opposed to mobile phones. 

\system enables a distributed data collection experience, with users having agency to choose how to take photos with their own tablet and position the camera and objects they are photographing. We imagined that this distributed experience could support the emergence of individual differences in data collection strategies, which may ultimately support groups diversifying their dataset (DDP1).

Image data collected in \system are stored and synchornized using private cloud-based data storage, with image data accessible only to users within a shared project.

\subsubsection{Train a Model}
Models are trained on device using the Create ML API \cite{createml}, with model training taking approximately 5-10 seconds for datasets of a couple thousand images (though larger datasets are supported). We minimized training time in an effort to reduce friction for users iterating on their model and datasets.  The trained model is stored locally and not synced across devices, as we imagined users might want to test the model in different states, such as by comparing model performance before and after modifying a dataset by adding more images.

\subsubsection{Evaluate Model Performance}
\begin{figure}[htb]
  \centering
  \includegraphics[width=\linewidth]{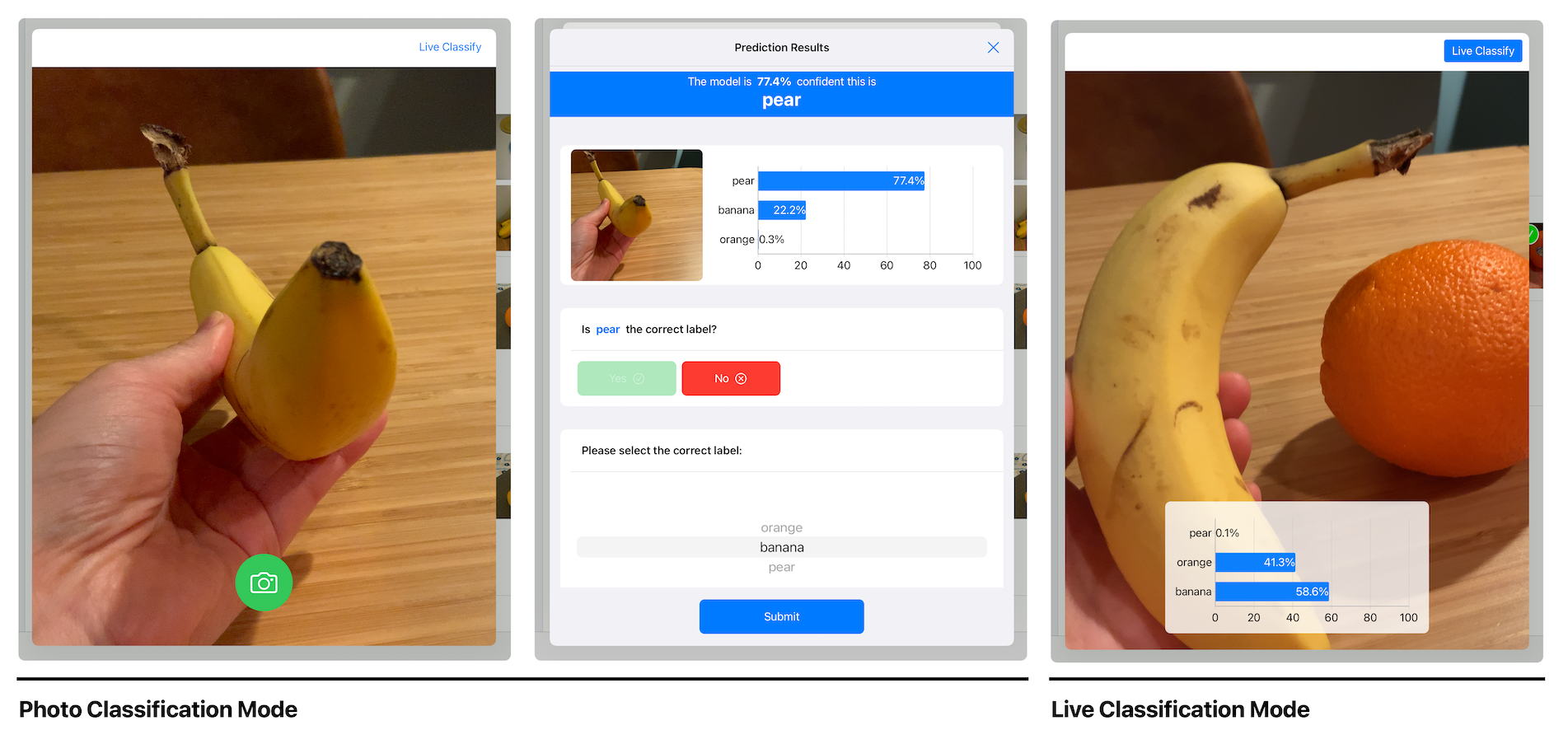}
  \caption{Classifying new data using the Photo Classification Mode (left) and Live Classification Mode (right).  In Photo Classification Mode, the user takes a photograph and can review the classification results. For misclassified data, they can relabel it with the correct label. In Live Classification, users can see an updating bar chart displaying relative confidence levels for each class in their model.}
  \Description{\todo{add description}}
  \label{fig:classification-mode}
\end{figure}

Immediately after a model is trained, a camera interface appears where users can test the model on new data.  They can take a photograph and view model results (Camera Classification Mode), or turn on real time classification (Live Classification Mode) where a live bar chart of confidence levels displays how the model is interpreting camera data in real-time similar to Google Teachable Machine.  When users capture images in Camera Classification Mode, they can indicate whether the model was correct in its prediction and provide the correct label if the data was misclassified, as shown in Figure \ref{fig:classification-mode}.  

\begin{figure}[h]
  \centering
  \includegraphics[width=\linewidth]{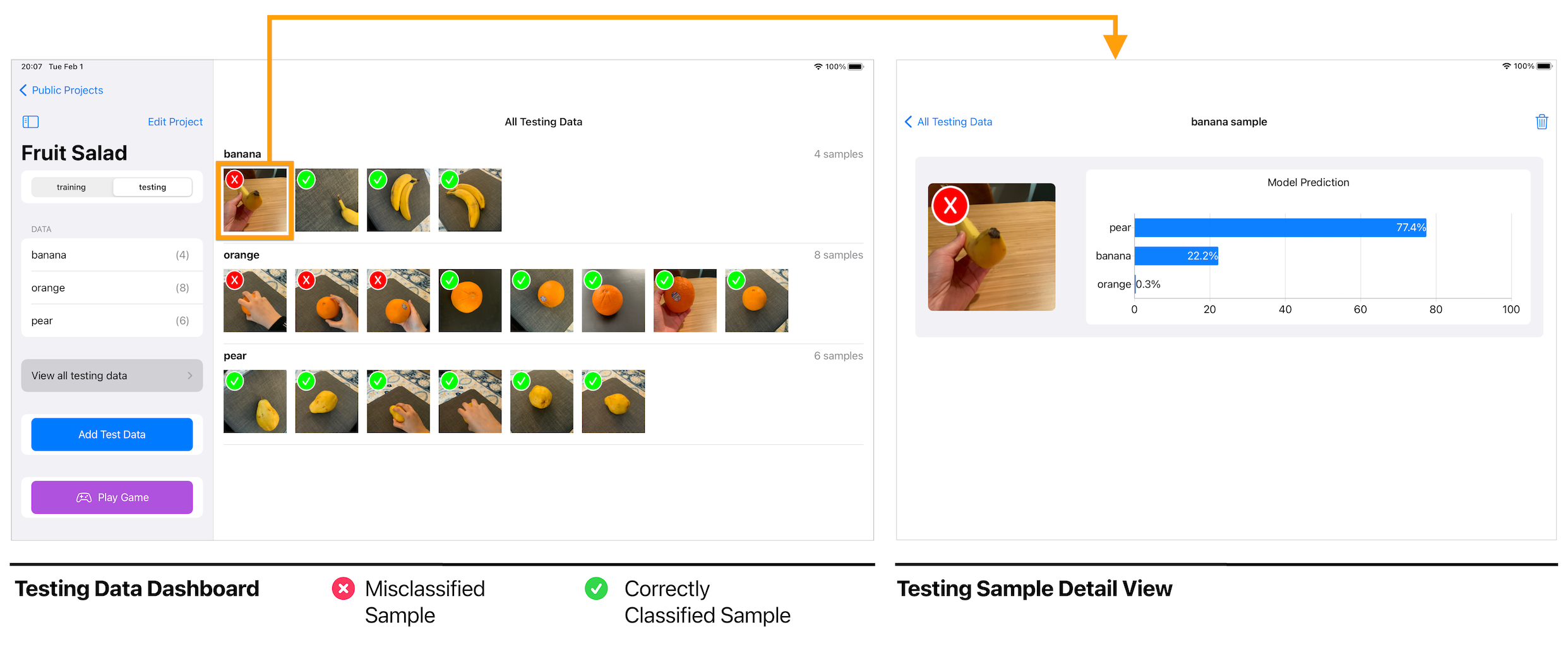}
  \caption{Testing mode interfaces.  Users can review collectively added test data, and classification results are based on the latest trained local model.  Tapping on a misclassified sample shows a bar chart of confidence levels to help users debug or improve model performance.}
  \Description{\todo{add description}}
  \label{fig:testing-mode}
\end{figure}

All images captured in the Camera Classification Mode are automatically saved as test data that can be reviewed in a Testing Data Dashboard (similar to the Training Data Dashboard).  The Testing Dashboard was designed to support student interpretation of model performance (DDP2). As shown in Figure \ref{fig:testing-mode}, the latest model classification results are shown, with green checkmarks indicating if the image was correctly classified and red x marks indicating that an image was misclassified.  The Testing Dashboard sorts misclassifications before correctly classified samples, encouraging users to identify differences in data that might lead to misclassification.  Further, if users revise their data (by adding or removing images), the testing dashboard always displays the latest result after a model is retrained– in this way, learners can check whether the total number of misclassifications for a given label changes over time.

\begin{figure}[htb]
  \centering
  \includegraphics[width=\linewidth]{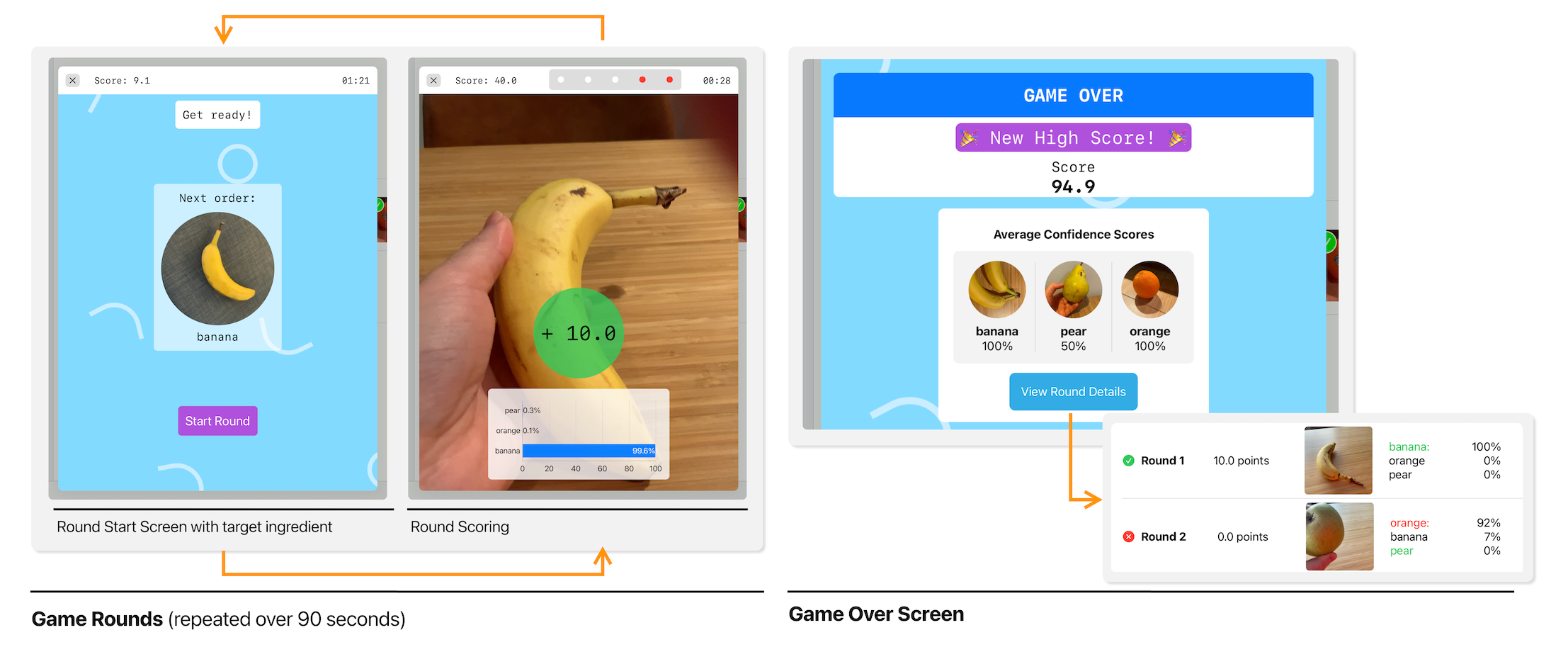}
  \caption{Model Evaluation Game. Users fulfill as many rounds as they can within a 90 second time limit.  Each round is 5 seconds, and the user gets a target object to show to the camera. The round score is calculated based on the confidence level of the image classified at the end of each round.}
  \Description{\todo{add description}}
  \label{fig:game-ui}
\end{figure}

We also created an in-app game as another way to support testing in a playful, structured format, as displayed in Figure \ref{fig:game-ui}.  The game consists of multiple rounds in which a target label is given, and the player must show that item to the camera.  They score points in each round based on the confidence level with which the model identifies the object (for example, an apple classified with a confidence level of 75\% would get 7.5 points).  The game consists of multiple rounds that the user completes within a 90 second time limit, designed so that users would be testing each label in their project multiple times.  At the end of the game, the user can see their cumulative score, a high score that serves as a proxy for monitoring whether a model has improved or not after iterating on the data, and details of the individual rounds to review misclassified items (DDP2).

\subsubsection{Iterating on a model}
At any point, users can revise their training and testing data by adding or removing images or adding or removing labels.  After making changes, they can retrain their model and test its performance to see if it has improved. The app uses a flat permission structure, with all users having the same abilities to add and delete data throughout the model building process.

\subsection{Starter App}
\label{sec:starter-app}

To support users applying models they create in \system, we developed a companion mobile starter app designed for beginners to program their own ML-powered apps. Users first export their ML model from \system, and then edit and write code in the starter app (using the Swift programming language) to build a real-time classification experience using camera input on mobile devices. 

With the starter app, users design and specify what information should appear when an item is classified by the camera. The underlying implementation (including the consumption of an ML model, instantiation of the camera, and real-time inference from live camera input) are handled under the hood to reduce development time and knowledge required to build an ML-powered app. 

\begin{figure}[h]
  \centering
  \includegraphics[width=1.0\linewidth]{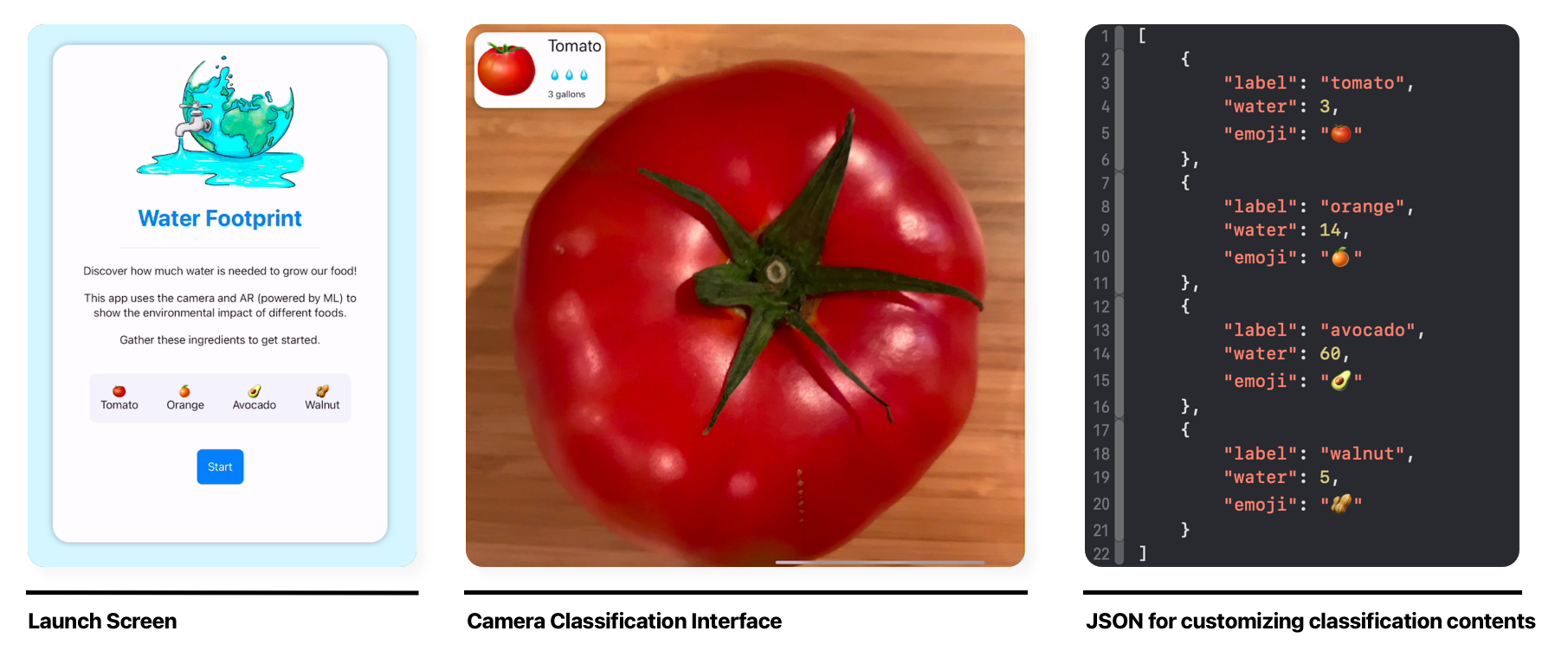}
  \caption{Starter app for building custom ML-powered app using models built in \system.  The app includes a customizable launch screen, where the purpose of the app and items the app can classify can be described.  The app then launches into a camera interface, where the information displayed when an item is classified can be customized. The content that appears in the camera view is defined by a JSON file; in this example, when the model returns a classification result of "tomato," the app displays a number ("water") for the gallons of water needed to grow a tomato and a corresponding tomato emoji ("emoji").}
  \Description{\todo{add description}}
  \label{fig:starter-app}
\end{figure}

The default experience of the starter app is displayed in Figure \ref{fig:starter-app}, where users can customize a Launch Screen describing the purpose of the app and the labels the classifier can identify. The app then exposes a camera interface for live classification, and users can specify the information they want to appear in the UI for the top classification result (in the example below, that a tomato requires 3 gallons of water to mature).  The specific properties to be surfaced in the camera classification interface is handled by a user-edited JSON file.

The source code for the starter app was provided for students to customize in the design of their own ML-powered mobile applications.

\section{Methods}

Our research team collaborated with the non-profit \nonprofitfull on the design and implementation of 2-week ML Summer Camps for young women and gender-expansive youth at the secondary education level. During these camps, students used \system in a variety of hands-on model-building activities, working in groups to develop final projects involving the design and development of custom ML models and apps. 

In this section, we first provide an overview of \nonprofitfull and the creation of the \campname design and structure. We then describe \nonprofitfull's recruitment strategy and how our research team recruited camp participants for our research study. As not all camp participants were part of our research study, we use \textit{student} to refer to anyone enrolled in the camp, and \textit{participant} to refer specifically to students that consented to participate in research. Finally, we describe our data collection and analytical approach to understand how participants used \system to support their understanding of ML data design practices. 

\subsection{\nonprofitfull and Creation of \campname}
\nonprofitfull is a non-profit that offers in-person and virtual summer camps for girls and gender expansive youth ages 13-18 in web development, mobile development, and data science in over a dozen cities across the United States, with 10,000 alumni as of Summer 2022. Camps are offered free of cost to all participants and do not require any prior computer science experience. \nonprofit's central mission is to empower voices that are typically underrepresented in technology and so the organization makes special effort to recruit diverse youth, with 82\% of program alumni identifying as people of color and 50\% qualifying for free and reduced lunch at school. (In the United States, free and reduced lunch qualification is calculated by multiplying Federal income poverty guidelines by 1.30 and 1.85 respectively \cite{Federal22}; free and reduced lunch is commonly reported as a measure for socio-economic status). 

In surveys with alumni, AI and ML was the top request for future camp offerings. Since \nonprofit did not have in-house expertise in AI or ML, \nonprofit partnered with our research team and others to design and pilot two in-person ML summer camps in Summer 2022: the first taking place in Seattle and the second held in New York City. Each camp was led by two \nonprofit instructors with prior experience teaching existing \nonprofit web development camps. In advance of the camp, \nonprofit's instructional team met regularly with our research team over several months to learn how to use \system and co-develop curricular activities. 

\subsection{\campname Structure and Activities}

During the \campname, students met Monday through Friday from 9am-5pm over two weeks for a total of 80 hours. In the first week, instruction was provided on a variety of AI/ML topics, covering the basics of what AI is, various types of ML systems and model types (lightly adapted from the MIT DAIly curriculum \cite{dailycurriculum}), and the dangers of bias in ML systems, all alongside team bonding activities and discussions around diversity in tech. Additionally, there were introductory lessons on the Swift programming language and using the Xcode developer environment to create iOS apps. Each student was provided with a laptop and a tablet to use for the entirety of the camp.

The \system app was introduced in two structured modules during the first week, each lasting between 3-4 hours and involving students working in groups of 3-4 people. In the first module, students built fruit classifiers using apples, oranges, grapefruits, and mangoes. In the second module, students built classifiers with items of their choice that they brought from home. For both modules, facilitators guided students through using \system's training and testing interfaces and the in-app game. Colored and patterned cloth swatches were provided for students to use as backdrops for their items so they could both capture data and test their models using diverse backgrounds. During these activities, the instructors emphasized that it was important to try to capture objects from a variety of different angles and perspectives to ensure that the model had more data about the appearance of objects in their classifier. Project groups for these two modules were randomly assigned.

The introductory \system modules were followed by a culminating final project, where students worked in new teams to build classifiers and apps on topics of their choice. 

\subsubsection{Final Projects}

The design prompt for the \campname final project was to ``Build an image classifier (\system) embedded in an app (Xcode and Swift) that addresses a topic of your choice.'' This open-ended project brief was designed so students had the space to tackle a problem of personal interest, curiosity, or relevance. 

Final project teams were formed by the instructional staff by the end of the first week of camp. The instructors grouped students based on their shared interests, which students indicated through a survey asking which topics they were most passionate about. Topics were selected from a list \nonprofit provided, including Animal Rights, Climate Change, and Creativity / Arts. Students were not bound to choosing a final project based on their initial shared interest; rather, these topics were used as a starting point for students to ideate on personally relevant projects themes.

The last week of camp was primarily unstructured work time for students to develop their projects. Groups pitched their final project ideas to the class and guest panelists on Monday of the second week, received feedback and suggestions, and then developed their ML models and custom over the next three days, with 9.5 hours total working time. Students built their models in \system by photographing items they brought from home to use during the camp. They then incorporated these models into custom iOS apps for creating a real-time image classifier using camera input (editing the starter app described in Section \ref{sec:starter-app}). Each team was assigned a Project Manager, a member of the \nonprofit instructional staff that monitored the team's progress and provided feedback and guidance as needed. 

Students documented their design process in digital design journals edited in Google Slides, with one project journal per group. The digital design journals provided templates for documenting model issues and revisions to their dataset, with space for adding screenshots of their \system projects and text descriptions of their process. We provided three different journal templates that students could duplicate when documenting their projects: \textit{Issues}, \textit{Wins}, and \textit{Changes}. The \textit{Issues} template invited participants to ``share any issues (unexpected outcomes or bugs) you identified when testing your model or app,'' providing space for describing the issue identified, ideas for what might be causing the issue, and what they tried in response to the issue (including whether or not their changes worked). The \textit{Wins} template was a space to ``celebrate breakthroughs and accomplishments your team made in developing your project'', with space for describing the accomplishment and sharing any insights that led to the breakthrough. Finally, the \textit{Changes} template could be used to ``capture if you decided to change directions in your final project'' and was intended to be used for major changes to models as opposed to smaller bug fixes.  In the \textit{Changes} template, space was provided for describing the change and their rationale for why the team made the change: ``What led to this decision, and how do you think it will affect the user experience of your app''?

On the final day of camp, all teams presented their final project to an audience of invited family and friends.

\subsection{Camp and Research Study Recruitment}

In the context of the analysis described in this paper, we focus specifically on the second \campname because for this camp, our team instrumented \system with event-based logging to enable more nuanced analysis of students' collaborative modeling practices (described more fully in Section \ref{sec:methods-data-collection}). In this section, we describe \nonprofit's recruitment strategy for the NYC camp and how our research team invited students to participate in our research study.

\nonprofit had a two-prong recruitment strategy for recruiting students to the NYC camp. First, they individually emailed alumni that had exhausted all 3 existing \nonprofit camps in NYC and invited them to attend the \campname pilot by answering three open-ended questions about their interest (including ``Why are you interested in AI or ML?''); of the 19 students that were invited, 10 applied and all were accepted in to the program. Their second approach was to recruit from the pool of students that applied to advertised \nonprofit camps (data science, mobile development, and web development); applications were scored by a member of \nonprofit based on several metrics (curiosity, community-minded, and motivation), and students were selected based on a variety of factors, including application scores, free lunch eligibility, and prior experience in \nonprofit camps.

On the first day of camp, a member of the research team introduced themselves and invited students to participate in our research study, outlining that participation was voluntary and required no additional work --- all students completed the same assignments as part of their participation in the camp regardless of study enrollment. Students that consented to participate returned signed consent forms from themselves and a parent or guardian (if the participant was under the age of 18), and study participants were grouped together throughout both weeks of camps.

\subsubsection{Study Participants}

A total of 26 female and non-binary identifying students participated in the NYC \campname, and 18 participants consented to take part in our research study (69\%). Study participants were between the ages of 15 and 18 years old and came from five different states from the East Coast and Midwest United States (with the highest representation from New Jersey and New York). Eleven (61\%) participants qualified for free or reduced lunch at school. Fourteen of the 18 participants self-reported their race and ethnicity (with 4 identifying as multiracial); 13 participants identified as Asian (Asian Indian, Chinese, Korean, or Other Asian), 1 identified as White, and 1 identified as American Indian. A majority of participants (78\%) were alumni of other \nonprofit summer camps, with 44\% having completed all three existing \nonprofit camps in web development, data science, and mobile development. 

\subsection{Data Collection}
\label{sec:methods-data-collection}
During the camps, 3-4 members of our research team were embedded in the classroom and captured participant experiences through classroom observations, notes, and audio recordings of participant dialogue; additionally, one member of the research team captured photos and videos of participants' interactions. During the final week, our research team shadowed 3 project teams to observe their end-to-end process of coming up with a project idea, training and iterating on their model with \system, and building their ML apps. After initial project pitches on Monday of the second week, the research team selected these 3 teams in an effort to represent a diversity of project topics and collaborative styles (based on observations we had of individual participant's working styles from the first week, with one team consisting of quieter participants and another with more vocal participants, for example).

To understand how individuals and teams collaboratively built models in \system, we instrumented the \system app to capture logs of actions taken on each tablet. These logs included each time an image was added to the model, whether that image was for training or testing the model, the label the image was associated with, and the raw image data for saved images. We also collected logs when the model was (re)trained, including the number of testing images in each label that were classified correctly, and when a user started the live classification interface or the game. Each of these events was recorded with an associated timestamp and an ID for the tablet being used. Along with log data, we captured screenshots of each \system project's training and testing dataset dashboards twice a day on Tuesday, Wednesday, and Thursday of the final week, at noon and at the end of the day. This enabled us to assess \textit{how} the number of training and testing images might have changed, as well as to qualitatively analyze features of deleted images (as we did not retain deleted image data in our logs to preserve user privacy).

Each project team had the option to grant the research team access to their \system projects, which included all images they added to their dataset, and all study participants granted permission. The image data captured in \system was stored in private cloud-based datastores and were only accessible to members of the research team and the participants within each project team. We also collected copies of participant artifacts (including design journals, presentation decks, and Xcode projects). Participant presentations were both video and audio recorded for our analysis. 

Throughout both weeks of the camp, participants filled out daily surveys about their experience, including a question designed to measure their confidence engaging in dataset design practices, which asked participants to rate how strongly they agreed with the statement "I can demonstrate how data can influence model performance" on a scale from 1 (strongly disagree) to 5 (strongly agree).The instructors allocated 10 minutes at the end of each day of camp for students to fill out the daily surveys. Daily surveys are part of the \nonprofit camp experience, regardless of participation in the research study; as most participants in the AIML camps had prior experience with \nonprofit summer camps, they were likely accustomed to the daily survey format.  

At the conclusion of the camp, we invited all study participants to 2-hour virtual debrief sessions using a semi-structured interview protocol. We asked participants for feedback about their overall camp experience and their use of \system specifically. All 18 study participants attended the post-camp debriefs, which were held 2 weeks after the conclusion of the camp, and received a \$50 giftcard for their time. To accommodate all participants' availability, we held two separate debrief sessions using the same format and facilitated by 3-4 members of our research team. 

Each debrief session began with a group discussion and reflection about the overall camp experience, followed by smaller breakout rooms with 3-4 participants each, where participants reflected on their final projects. Breakout groups were composed of participants from different final project teams in an effort to 1) encourage participants to compare and contrast approaches taken on by other teams, and 2) enable our research team to investigate descriptions of design process from each participant in a team to learn how individuals contributed to group efforts. We captured video and audio recordings of these virtual debriefs for our analysis of participants' reflections on their experiences. 

\subsection{Analysis}

We take a constructivist stance toward studying and designing for learning; that is, we theoretically conceptualize learning as knowledge building. Learning happens through activities, wherein people activate and draw upon prior understandings to make meanings of information around them, and then build new understandings when interactions with their environment challenge the completeness or utility of prior understandings \cite[pp. 8--14]{bransford2000people}\cite{margulieux2019learning}. This theoretical stance necessitates attention to the processes through which people learn, including what they attend to, and how they make sense of their observations, explain their understandings and consequent actions, and change strategies as their understandings change over time. We attend to process through documentation and analysis of the modeling actions that participants take as they use Co-ML, especially moments when they make sense of model performance, relate that performance to characteristics of training and testing dataset compositions, and coordinate their activity with one another, which includes justifying proposed actions with their interpretations of their observations. 

Because we take a constructivist stance, the best evidence of learning is in participant-driven ideas and implementation of dataset design practices; as a result, our analysis focuses on participants' final projects, where teams chose what apps to develop, what data to use, and how to improve their models over 9.5 hours of working time across 3 days. Our research team anticipated that we would see the most variation in debugging scenarios and approaches during this time as a result of teams working on distinct project ideas.

In this section, we describe our approach to analyzing data collected during the \campname. Our research team met regularly over the course of 10 months to discuss the data and reach consensus on our interpretations of how DDPs were represented in participants' modeling practices.

\subsubsection{Data Cleaning \& Preparation}

Before our team began analysis, we cleaned and prepared the log data. We reviewed all participant-collected image data in \system and removed images containing any personally-identifiable information to preserve participant and non-participant privacy. Criteria to remove images were created in a discussion of four authors, and then applied to a random sample of images. After another round of discussion, the criteria were refined and then a single author flagged images for removal. Flagged images were reviewed by all four authors who developed the criteria. In total, of the 6,756 total images students captured in their final projects, 4.2\% of images were removed. Removed images included images in which non-study participants were inadvertently captured or personally identifiable information like full names were visible.

In addition, we cleaned the log data so that logs only represented actions completed in \system by each study participant (as opposed to actions in \system taken by the research team during our data collection throughout the camp). 

\begin{figure}[h]
  \centering
  \includegraphics[width=\linewidth]{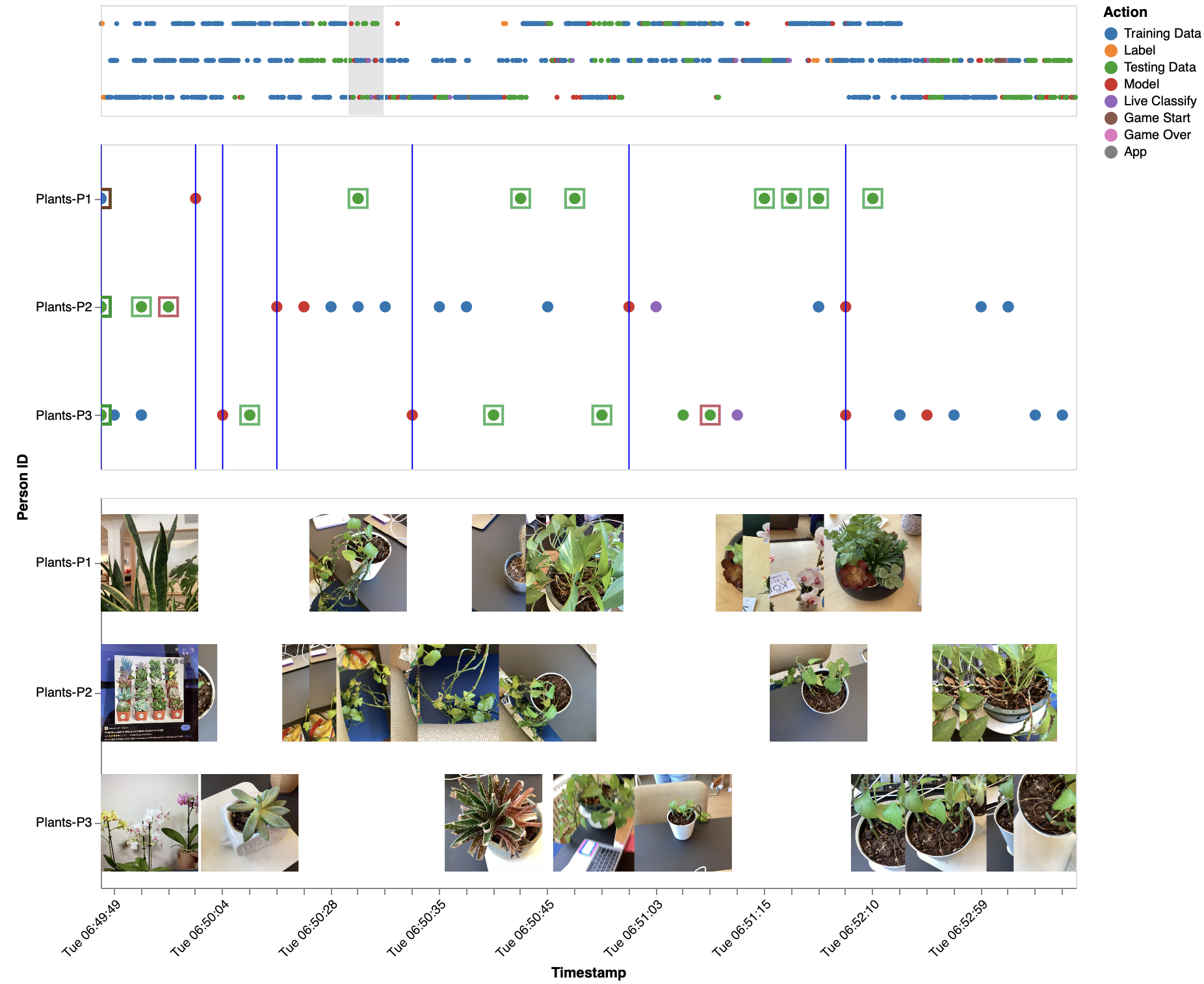}
  \caption{The visualization tool we created to view timelines of team activity. The top chart includes one row per team member and allows selection of smaller time intervals, which are then displayed in the lower two charts. The middle chart displays one row per team member, with a colored circle to indicate the action taken in \system and blue lines indicating when the model was retrained on a device. The bottom chart displays thumbnails of images added by each individual. The middle and bottom charts are aligned on a single horizontal axis representing time, allowing researchers to compare actions taken and images added across individuals through time.}
  \Description{}
  \label{fig:log-viz}
\end{figure}

We then created visualizations from our \system log data to construct timelines of team activity. Using our visualization tool, we could trace 1) an individual's model building process and what features of the app they were using (e.g., if they added training images of one label, trained the model, and then tested the model), and 2) what their team members were doing in parallel (e.g., we could see if two team members were adding data to the same label simultaneously). Figure \ref{fig:log-viz} shows a screenshot of this tool on data from one of the project teams in our study, displaying how our research team could inspect data added by individuals over time. Hovering over any image displays a tooltip indicating whether the image was added to the training or testing dataset; for testing images, we could also inspect whether the image was classified correctly and what the top predicted label was from the model.

Finally, we transcribed audio from three different sources: recordings of conversations between team members while they worked on their final projects, the final presentation each team gave, and the post-camp debrief interviews. 

\subsubsection{Description of Modeling Actions}

One of our first analysis priorities was to understand how frequently students used various features of \system. This let us see how participants and teams actually constructed the models in their final projects. To do this, we primarily looked to the log data. For each app feature we calculated the frequency of use for both participants and teams. For example, we calculated how many times members of a particular team trained their model and how many times they played the game. We also calculated the size of training and testing datasets for each group. Finally, we analyzed screenshots of \system that students incorporated into their project journals and final presentations to provide evidence about model performance. While these screenshots are not exhaustive of all features they used in \system, they provide an indication of which parts of the \system interface students felt supported their descriptions of how well their model was working.

\subsubsection{Identifying Data Diversity Strategies from \system Datasets} 

To answer our question of how participants considered dataset diversity (DDP1), we examined the image data from all team \system projects using a grounded-theory approach \cite{glaser2017discovery, strauss1990basics, charmaz2000grounded}. We began with inductive open coding of the corpus of images from final project \system datasets to categorize different strategies for incorporating data diversity (such as if teams took photographs of objects from different angles, with different backgrounds, or captured multiple potential form factors of an object). The open coding was performed by the lead author, then collectively discussed with three other members of the research team to identify additional codes and to cluster codes based on similarity. Once our group was in agreement on all four overarching data diversity strategies identified in the data, we then reviewed the image data again to deductively code each team's dataset, leading to frequency counts of how many teams utilized each data diversity strategy in constructing their datasets.

\subsubsection{Identification of Collaborative Modeling Moments} 
To understand how participants engaged with each DDP while collaboratively building models with \system, our research team identified model debugging moments from each team through a multi-step and multi-source process. Analysis of these debugging scenarios ultimately helped us construct in-depth case studies of how \system and peer-to-peer interactions helped participants enact DDPs in refining their models.

First, we reviewed transcripts of team final presentations and final presentation slide decks, where participants provided descriptions of challenges they worked through when implementing their models and mobile applications. The presentation decks often included screenshots from \system that complemented their verbal descriptions; for example, when describing how they resolved a misclassification scenario, some teams included screenshots of the testing dashboard in \system showing how many of their samples were classified incorrectly before and after revising their datasets. 

Second, we reviewed transcripts of post-camp debriefs, where students described these debugging scenarios in more detail in conversation with our facilitators. As participants were placed in mixed breakout groups containing participants other than their final project teammates, we compiled per-team documents containing statements from each team member to more fully characterize each individual's role in the debugging scenario. 

Third, we examined how participants described model issues in their project journals, including what screenshots of the \system app they used to supplement descriptions of model issues, ideas of what they thought might be causing the issues, and what they tried in response to their ideas (and whether or not those changes improved the model). 

Fourth, we reviewed our visualization of the log data to reconstruct participants' actions in \system during these debugging moments. To review the logs, we had to first determine approximately when these debugging scenarios took place. If a member of our research team observed the team directly, we reviewed dated observation notes to determine approximately when the scenario occurred. If needed, we also examined screenshots of \system from project journals or final presentation decks to determine, using our logs, when a particular image sample was created. For example, if a team captured an example of a misclassified image in their design journal, we could find the image in our logs and determine when it was captured. We then selectively reviewed our visualizations around the identified time when the debugging moment occurred to corroborate participants' reflections on their process from our interviews and their design journals.

Finally, our fifth step was to supplement log timelines with select transcription of participant dialogue from our audio recordings captured around the time of the debugging scenario. This let us triangulate using discourse between participants while debugging, logged actions they performed in the app, and reflections participants had on their experience.

We then deductively coded the debugging moments based on the DDPs we identified from the literature (Section \ref{sec:dataset-design-practices}). While each debugging moment might have encompassed multiple DDPs, we ultimately chose one in-depth case study for each DDP to best illustrate what the practice looked like as participants used \system.

\section{Results}
\label{sec-results}
We begin by providing an overview of the final projects participants built, including the problems they designed for and summary statistics about their datasets and model iteration. Next, we illustrate the ways collaboration shaped team understanding of the role of data in ML by considering learning with respect to each of the four dataset design practices from Section \ref{sec:dataset-design-practices}.

\subsection{Final Projects Overview}

Participants worked in teams of three to create custom ML-powered mobile apps centered on topics of their choice, ultimately addressing substantive issues around racial inclusion, consumer responsibility, and sustainability across diverse domains like food, fashion, and health. Table \ref{tab:final-projects-summary} shares a summary of the projects created by the 6 teams along with each of their models’ corresponding labels. The diversity of ideas and applications participants built suggests that \system and the companion starter app flexibly supported a range of personally-relevant projects, ultimately inspiring participants to consider the multitude of ways ML can impact their lives. As one participant from Team Fashion (Fashion-P3) described: ``All of our projects were about an important issue in society... We learned what AI and ML is and what it does, and more importantly, we learned how to take those skills and apply it to address important issues in the world.'' 

\begin{table}[h]
    \centering
    \begin{tabular}{ l p{5cm} p{5cm} }
    \toprule
    Team Name & Project Description & Labels \\ \hline
    Plants & Classifying house plants and providing information on how to take care of them  & cactus, succulents, pothos, orchids, monstera, snake plant \\ \hline
    Fashion & Identifying fashion labels from clothing tags and providing information about the brand's environmental impact & CHNGE, F21, H\&M, Patagonia, Zara \\ \hline
    Donations & Identifying categorizes of items that can be donated and where to donate them  & Technology, stationary, books \\ \hline
    Nutrition & Providing nutritional information about food items & Orange, cucumber, mint tea, mayonnaise, poptarts, juice \\ \hline
    Foodwaste & Recommending recipes that use common leftover ingredients & Avocado, onion, orange, apple, tomato, sliced avocado, sliced tomato, sliced onion \\ \hline
    Makeup & Revealing how inclusive different makeup brands are based on the range of skintones they support & Glossier, Covergirl, Clinique, Fenty, Neutrogena \\ 
    \bottomrule
    \end{tabular}
    \caption{Summary of team final projects}
    \label{tab:final-projects-summary}
\end{table}

\begin{figure}[h]
  \centering
  \includegraphics[width=\linewidth]{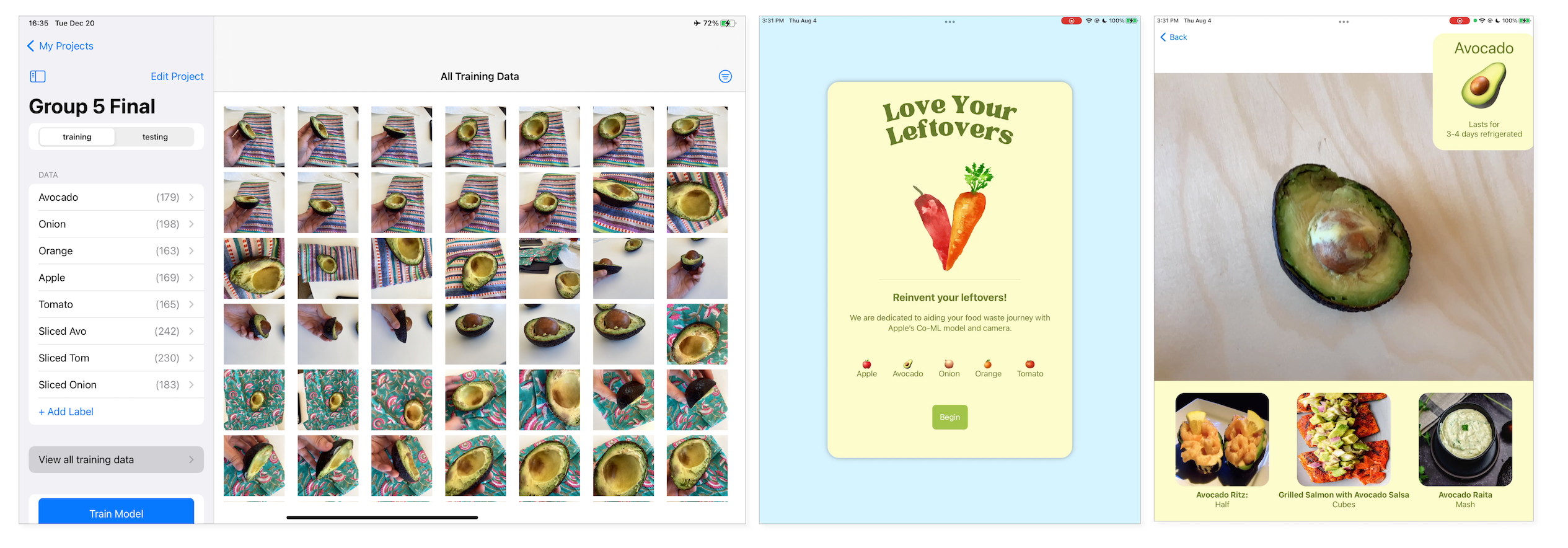}
  \caption{Final project by Team Foodwaste, who built an app to recommend recipes for leftover ingredients. Their training data in \system (left) incorporated images of produce on different surfaces, and their app (right) used their image classifier to recommended recipes based on identified ingredients.}
  \Description{}
  \label{fig:ll-final-project}
\end{figure}

Figure \ref{fig:ll-final-project} presents an example final project from Team Foodwaste, who created an app centered around reducing food waste by recommending how to use leftover produce. To create this app, they constructed a classifier in \system to identify produce like tomatoes and onions and programmed their starter app to suggest recipes for identified ingredients.

Each project had between 3-7 labels, and their model's data consisted of participant-collected images of objects they photographed during camp. Objects were brought from home (such as houseplants and pieces of clothing) or available at the camp (like laptops, books, plants, and snacks). Additionally, half of the teams (Teams Plants, Fashion, and Makeup) incorporated photos of image search results from the Internet to supplement their datasets.

Teams had an average of 1,000 images in their training data, with a range of 399-1,609 images (SD=475). In post-camp debriefs, participants noted that this dataset size was possible because of the distributed data collection experience in \system, where multiple people could contribute data simultaneously. Student Foodwaste-P1 shared, ``I think the process of getting data and taking the photos became so much more efficient – the fact that we were on different iPads. If it was just one iPad that we were just passing around, it would probably take us so long just to get all of those photos together.''  Projects had on average 220 testing images (SD=53).  While these datasets are smaller than production-level ML models (which might have hundreds of thousands to millions of images), the underlying architecture of \system leveraged transfer learning to enable users to train models on smaller datasets that were still on average 80\% accurate to their test datasets as shown in Table \ref{tab:project-data-results}. Accuracy was calculated by weighting each label’s accuracy by the number of images in the testing set, then summing those weighted accuracies to get an overall picture of how well the model was performing on the testing data.  

\begin{table}[h]
    \centering
    \begin{tabular}{ l  c  c  c  c }
    \toprule
    Team & Training Images & Testing Images & Accuracy \\ \hline
    Plants & 657 & 174 & 0.93  \\ \hline
    Fashion & 399 & 231 & 0.51 \\ \hline
    Donations & 985 & 203 & 0.89 \\ \hline
    Nutrition & 1,521  & 302 & 0.82 \\ \hline    
    Foodwaste & 1,187 & 157 & 0.89  \\ \hline
    Makeup & 1,609 & 253 & 0.86  \\ 
    \bottomrule
    \end{tabular}
    \caption{\label{tab:project-data-results} Descriptive statistics of final project models and their datasets}
\end{table}

In the following sections, we describe the ways in which each dataset design practice was enacted, with participants collaboratively designing their datasets and interpreting model evaluation results to refine their data and corresponding models.

\subsection{DDP1: Incorporating Dataset Diversity}
\label{sec:results-ddp1}

Participants considered dataset diversity as it related to their envisioned user experiences for their apps. Our analysis of \system datasets identified four types of dataset diversity strategies: \textbf{perspective}, \textbf{contexts}, and \textbf{states} relate to diversity on a single object, while \textbf{types} involved variation on a class of objects. Table \ref{tab:data-diversity} defines each of these strategies and provides examples from team projects.

\begin{table}[h]
    \centering
    \begin{tabular}{ l p{4cm} p{5cm} }
    \toprule
    Strategy & Definition & Examples \\ \hline
    \textbf{Perspectives} & Varying viewpoints on a single object  & Rotated object \newline \newline Crop and size of object \\ \hline
    \textbf{Contexts} & Varying the background around an object & Color or pattern of background \newline \newline Presence of occluding objects (e.g., hand holding item) \\ \hline
    \textbf{States} & Varying the form factor or condition of an object & Items opened or closed \newline \newline Ingredients slice or whole \newline \newline Condition of an object (e.g., health or sick plants) \\ \hline
    \textbf{Types} & Varying the types of objects within a class & Distinct form factors (e.g., over-the-ear-headphones versus in-ear earbuds) \\ 
    \bottomrule
    \end{tabular}
    \caption{\label{tab:data-diversity} Data diversity strategies represented in team projects}
\end{table}

\textbf{Perspectives} on an object included capturing multiple angles or camera zoom levels on a single object and was represented in all projects. \textbf{Contexts} on an object (also represented in all projects) involved photographing objects on different colored backgrounds or having occluding objects like hands represented in part of the data. \textbf{States} of an object encompassed multiple form factors or conditions, such as whether a fruit was cut or whole or if a plant was health or dying; states were represented in 4 of the 6 projects. \textbf{Types} incorporated different kinds objects to represent a single label and was a strategy used by just two teams due to the specific use case for their apps. For Team Makeup, variation within a class involved capturing both 2D product images and physical 3D products to account for shoppers who might use their model in an physical retail shopping experience or online.  For Team Donations, each label covered multiple types of objects; for example, under the label ``Technology,'' they took photos of phones, laptops, and earbuds. For all teams, multiple strategies were used in combination to diversify their datasets. 

While both training and testing data were largely captured in the same classroom, participants took measures to mimic different contexts by using patterned or colored fabrics (provided by the instructional team) as backdrops for their objects, or by testing with images from the Internet (which half the teams did). For example, Team Fashion wanted their model to identify brand labels on pieces of clothing; since they had a limited amount of clothing they could bring to the camp, they decided to expand their dataset by taking photos of web search results of clothing of different colors. Our debriefs revealed how participants had even more ideas for how they would like to add diversity to their project's data if they had the opportunity to work in a different environments, with participant Foodwaste-P3 describing how ``The space that we had [classroom] was very well lit, and I think that was something else we were trying to do get – darker lighting and more diversity in our dataset.''

Some data diversity strategies were discussed a priori within teams at the onset of their data collection process. Since Team Plants recognized that plants could be in many different rooms in a house, they focused on photographing plants in different settings such as on a ledge, in front of a window, or on a table (an example of a \textbf{context} data diversity strategy). 

Other strategies arose through the use of \system as participants reviewed their collective datasets and noticed gaps in how objects were represented. This process was aligned with our design goals for \system to help surface differences in how individuals capture data and test models, as well as enable learners to discuss and interpret those differences.  We illustrate this process using a vignette from Team Foodwaste in which the team expanded their data collection strategies based on monitoring differences between how individual photographed their objects.

\subsubsection{Vignette: Noticing Differences to Diversify Data}
\label{result:LL-case-study}

Team Foodwaste's project centered around identifying produce items, and their image data incorporated multiple data diversity strategies: \textbf{perspectives}, \textbf{contexts}, and \textbf{states}. At the start, the team discussed capturing the produce items on different backgrounds to simulate kitchen countertops of varying colors so the model would be able to work well regardless of the surface an ingredient was on (an example of \textbf{contexts}); one participant captured images of the ingredients on a black coffee table, another on a wooden surface, and the third on a white table.  

\begin{figure}[h]
  \centering
  \includegraphics[width=\linewidth]{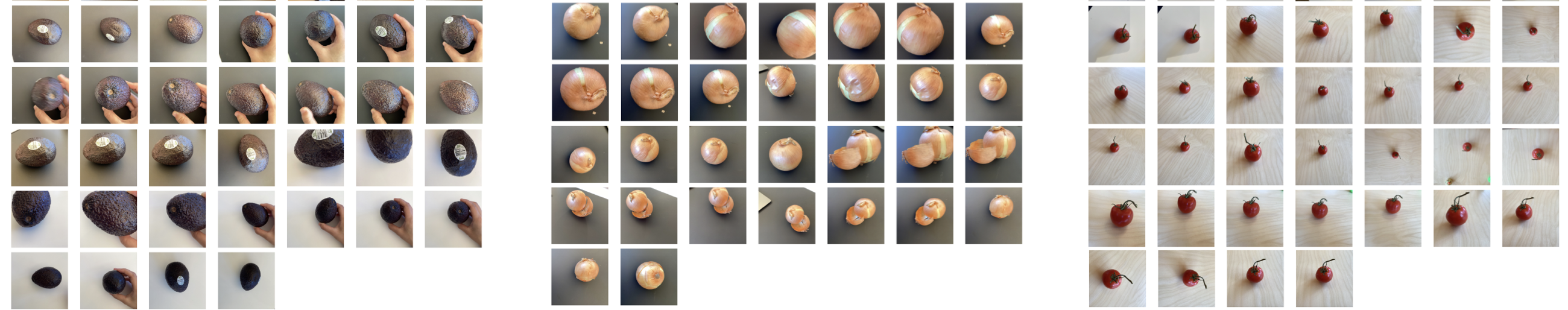}
  \caption{Initial training images from team members in Team Foodwaste, where avocado training images are the only on es that initially have the presence of a hand.}
  \Description{}
  \label{fig:ll-training-data}
\end{figure}

In contrast, other data collection strategies emerged through the modeling process. In her initial set of training images, participant Foodwaste-P3 captured photos of a whole avocado, with some training images showing her hand holding the avocado, and some capturing the avocado by itself. At the same time, her two teammates were capturing photos of onions and tomatoes, respectively, without any hands visible as shown in Figure \ref{fig:ll-training-data}.

After reviewing their collected images using the training data dashboard in \system, Foodwaste-P3 asked her teammates if they were putting hands in their photos, at which point the other two team members realized that they did not. As a result, they began to add images of both the isolated ingredient and the ingredients held by their hands. This was important as they wanted their model to work regardless of how the user presented the object to the camera. In our post-camp debriefs, Foodwaste-P3 reflected on this experience, sharing that while each team member was adding training data, she would periodically review the training dataset to monitor differences in their strategies:

\begin{quote}
I was going through photos that my other teammates had taken. And then I was trying to make sure my data was similar to theirs, or had the same diversity those did. And so that was when we started to realize this [not having hands in the training images] was an issue, but it wasn't something we had discussed. So, I think it was helpful that we could see the kind of training data that other people were taking. And then talk about it and improve our model that way.
\end{quote}

This same participant then went on to describe how perspective-taking worked in both directions --- she brought attention to differences in data collection strategies to help her teammates, and she also learned from her teammates’ perspectives:

\begin{quote}
Personally when I was taking photos, I was very limited in the kind of photos I was taking. I was taking very close up and certainly specific angles, but my other teammates were taking them farther away or with other distracting objects in the background. Which is, good, I think to diversify our dataset.
\end{quote}

Thus, we observed that this team was able to discover ways of diversifying their dataset by examining differences in individual data collection strategies with \system, aligned with our goal of enabling multiple perspectives to emerge from a distributed data collection experience. 

\subsubsection{Diversity in Response to Model Failure}

Incorporating dataset diversity was also driven by identifying failure cases of the model. Team Nutrition found that because only one team member initially collected training data in which a hand was visible for the orange label, any item held by hands was incorrectly classified as an orange, leading the team to add more images of hands holding objects to each of their labels. Team Makeup reflected on their strategies of adding diverse \textbf{perspectives} and \textbf{contexts}, as described by student Makeup-P1: “We did that [resolved misclassifications] by adding more training data of misclassified products while being more inclusive with the images that were inputed. So we took pictures from different angles or different lighting and we were especially conscious of the backgrounds of the new images. By incorporating new backgrounds, we were diversifying our dataset.” Dataset diversity thus was seen as a way both to account for different intended use cases and contexts, but also as a way to improve classification accuracy by accounting for dataset gaps, which we describe more fully in the next section.

\subsection{DDP2: Evaluating Model Performance and its Relationship to Data}
\label{sec:results-ddp2}

One of the affordances of the \system software is that models train in a few seconds on tablets, allowing teams to quickly experiment with their models and the data they were trained on. All teams took advantage of this affordance by retraining their models between 17 and 75 times, with groups retraining their models an average of 40 times (SD=20), and individuals retraining on average 13 times. (Note that with \system, each person needs to retrain on their iPad to have the most up-to-date model). The high frequency of retraining indicates that participants were able to iteratively test and continually refine their datasets. 

Participants leveraged several different features of \system when evaluating model performance between model retrainings. Our logs revealed that all teams utilized the testing dashboard for reviewing test images, as well as the live classification feature for viewing model results in real time, the latter of which was invoked an average of 11 times for each team (SD=5).  

Four of the six teams (all except Teams Nutrition and Donations) used the game to test their models. Teams Plants and Fashion played the game one time each, while Team Makeup played 4 times and Team Foodwaste played 11 times. Participants described how the game enabled them to monitor their progress. For example, participant Fashion-P1 shared, “I really enjoyed the testing game because it allowed us to see our progress and see how we've improved and what we added that improved the model's accuracy. I feel like we could test out some piece of data and be able to see how the [model] fared.”  We observed that Team Foodwaste developed a system for continually refining their model using the game; periodically, after their team had collected more data, each team member would play the game and review the Game Over screen to assess which labels were performing better and in what instances the model failed. Based on these results, they would decide which labels to add or delete data from. Participant Foodwaste-P1 described how their goal was to keep improving their model until all members of her team were able to get ``100\% prediction scores'' in the game.

\begin{figure}[h]
  \centering
  \includegraphics[width=\linewidth]{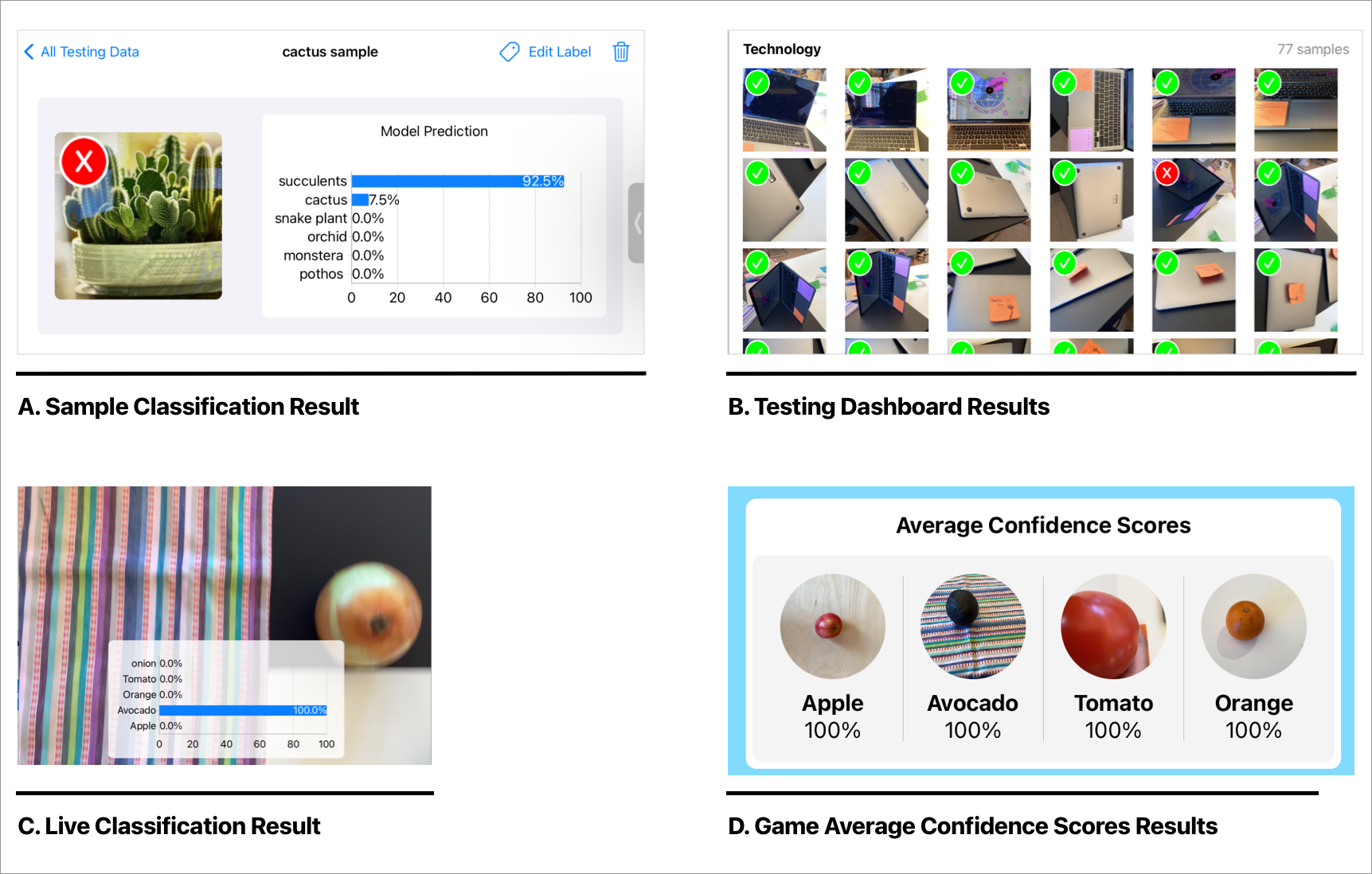}
  \caption{Screenshots participants incorporated into their project journals and presentation decks that incorporated UI screenshots from \system, including sample classification results, the testing dashboard, live classification results, and game feedback.}
  \Description{}
  \label{fig:project-journal-examples}
\end{figure}

Our analysis of team project documentation revealed which features of the \system app participants documented most frequently to support descriptions about how they iterated on their models. Figure \ref{fig:project-journal-examples} displays examples of screenshots of \system from team project documentation, which fell into one of four categories: 1) a sample classification result (showing the result of a single test image); 2) the testing dashboard (displaying a set of testing images and their classification results); 3) the live classification interface (where model results are shown in real time); and 4) the average confidence scores achieved while playing the game. The majority of teams (five out of six) incorporated screenshots of classification results for a single sample, while half of the teams included images of the testing dashboard. The live classification interface and game average confidence scores were incorporated into the design journal of a single team (Team Foodwaste). 

The combination of frequent model retrainings alongside repeated use of the testing dashboard suggests the value of integrating test datasets into a novice ML experience. And \system further enabled this exploration through a collaborative experience where learners could debate interpretations of model failures and decide how to iterate on their data collectively. To illustrate the role of collaboration in evaluating models, we provide a vignette of project Team Plants, who wrestled with a misclassification for their houseplant classifier.

\subsubsection{Vignette: Collaborative Model Testing}
\label{result:PS-case-study}

\begin{figure}[h]
  \centering
  \includegraphics[width=\linewidth]{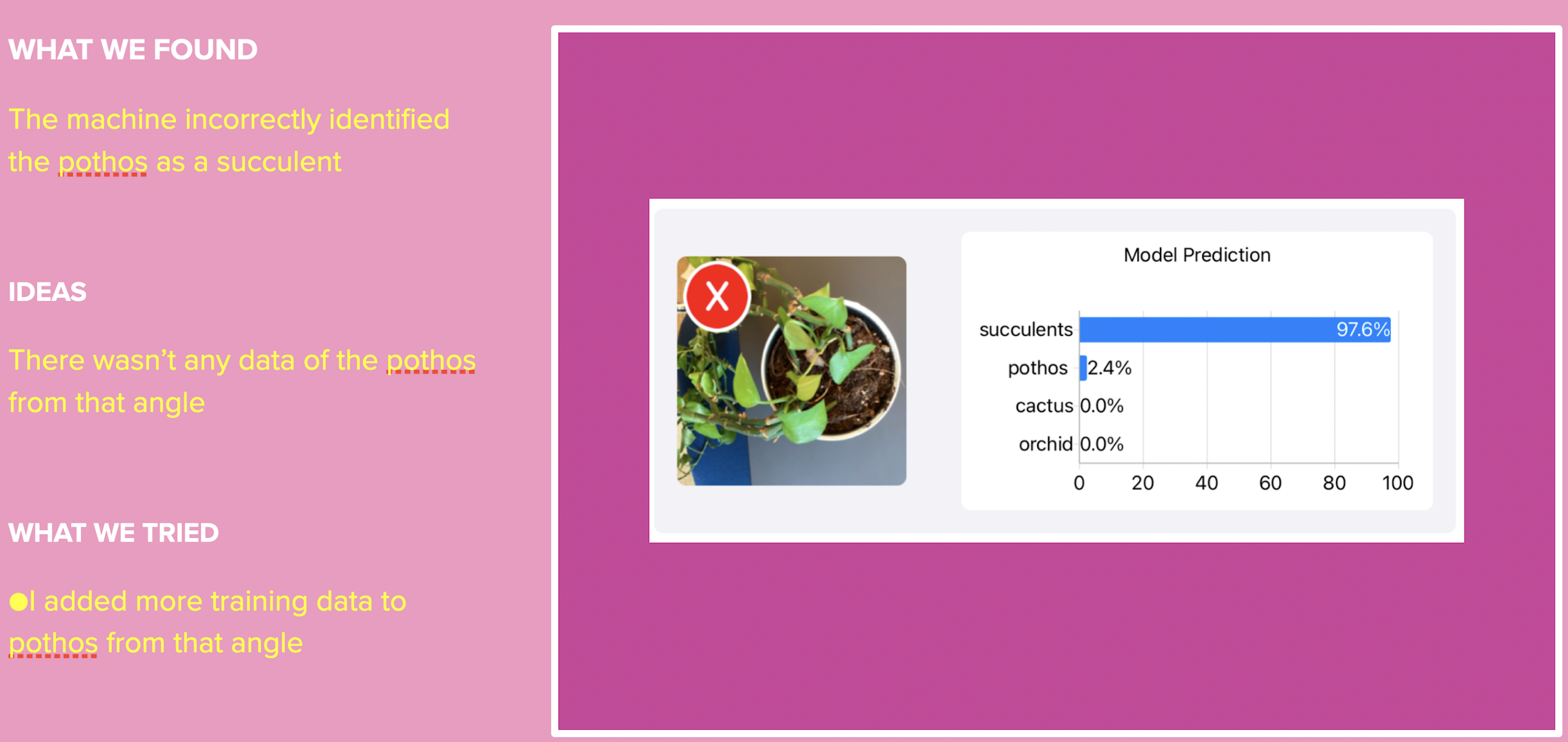}
  \caption{An excerpt from the Team Plants' design journal capturing their debugging of the Pothos and Succulent misclassification scenario}
  \Description{}
  \label{fig:ps-pothos-journal}
\end{figure}

Team Plants developed a final project centered around identifying houseplants and providing information on how to take care of them. When testing their model, each team member individually discovered that their Pothos plant was being misclassified as a Succulent, ultimately realizing that their training dataset lacked images of the Pothos plant shot from above as displayed in their project journal in Figure \ref{fig:ps-pothos-journal}.

 In the dialogue below, two members of Team Plants discussed the misclassification together:
 
\begin{itemize}
    \item  \textbf{Plants-P3}: When I’m doing it [testing the model] at this angle, [the model’s] saying the Pothos is a Succulent.
    \item  \textbf{Plants-P2}: Ok, got it.  I was doing the same thing [testing the model on Pothos].  Like, from the top angle, it [the model] kept getting [misclassifying the Pothos as] the succulent. I’m going to try to take some pictures of it.
    \item {[Both participants begin adding more training images of the Pothos plant]}
\end{itemize}

Through discussing the misclassified Pothos images, Plants-P2 and Plants-P3 decided to work together to add more images of the plant and discussed ideas for what might be causing the model failure:

\begin{itemize}
    \item  \textbf{Plants-P2}: I’m trying to take more photos of the [Pothos] stem because I guess that’s mainly near the top.  Or do you think it’s...because the succulent is so close to the soil, and that one [the misclassified pothos] is close to the soil maybe?  If that makes sense?
    \item  \textbf{Plants-P3}: Yeah.  Like, the soil [of both the Pothos and Succulent plants] looks similar.
    \item  \textbf{Plants-P2}: And maybe it’s [the model’s] using that to identify it, rather than the actual plant?  Cause this one’s [the Pothos], the soil’s kind of covered with the stem [of the plant], so I don’t know if it’s looking at that.
    \item  \textbf{Plants-P3}: Oh that’s true, yeah.
\end{itemize}

Through their discussion, both participants decided that the soil may be a confounding factor-–that the model is attending to the soil to incorrectly classify any images with soil as Succulent. Once they took additional photos of the Pothos plant with soil visible, they retrained their model and tested it on new data.

\begin{itemize}
    \item {[Plants-P3 tests the retrained model on an image of the Pothos.  The model classifies the Pothos plant correctly.  Next, she takes a photograph of a succulent, but the succulent is now misclassified as Pothos.]}
    \item  \textbf{Plants-P3}: I think I fixed it, but now it’s saying the Succulent is the Pothos.
    \item \textbf{Plants-P1}: Fix one thing, break the other [laughs]
\end{itemize}

While their refined model appeared to work better at identifying their previously misclassified Pothos plant, its performance for identifying Succulent was negatively impacted, leading the participants to recognize that the model needed to be more comprehensively tested. This is a common issue in ML practice, in which edits to a single label might affect model performance for other labels.

In this example, collaboration contributed to group awareness of the model failure, discussion and ideas for the cause of the misclassification, and coordinated data collection efforts, positively contributing to participants' discovery and reasoning about gaps in their dataset.

\subsection{DDP3: Balancing Datasets}

We designed \system to encourage the creation of balanced datasets by displaying sample counts throughout the interface, such as when users add images via the camera or review their datasets in the training dashboard. Our intention was for learners to use this information to identify if there were fewer samples for particular labels, and to ultimately create datasets with a similar number of samples for each label. Instead, we observed that while some teams initially set targets for equally distributed data collection across labels, teams ultimately took a more reactive approach, attending to specific labels based on model evaluation results. In fact, some groups intentionally created skewed datasets to improve performance for underperforming labels.

\begin{figure}[h]
  \centering
  \includegraphics[width=\linewidth]{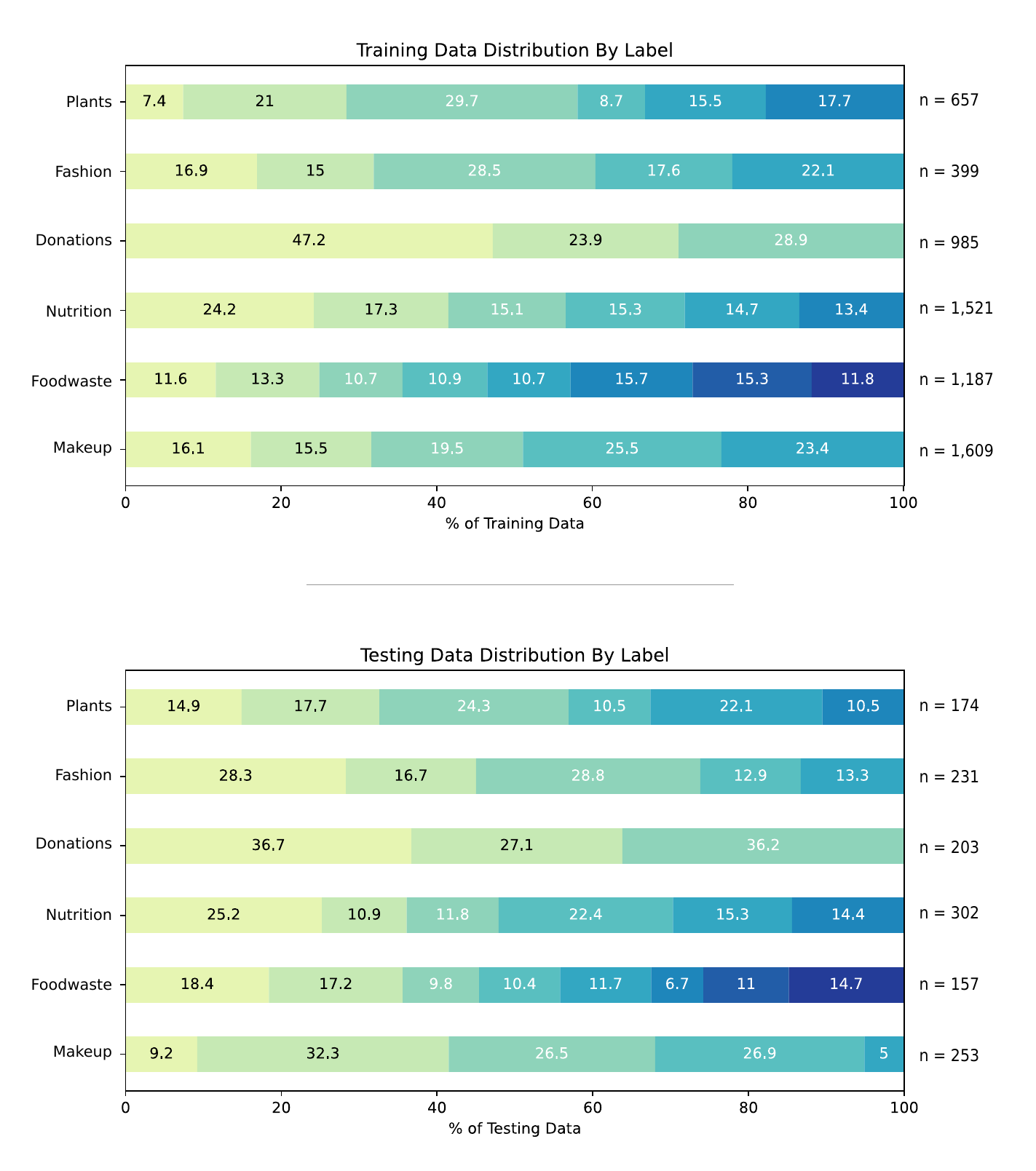}
  \caption{The distribution of training and testing images for each label in a given project. Each segment represents a label in a project, and the ordering of labels (left to right) matches the labels listed for each project in Table \ref{tab:final-projects-summary}}
  \Description{}
  \label{fig:training-testing-distribution}
\end{figure}

Figure \ref{fig:training-testing-distribution} reports the percent distribution of training and testing samples for each label within a project. First, these plots reveal differences in sampling among training data; for example, Team Plants had fewer Cactus (7.5\%) and Orchid (8.7\%) images relative to other labels in their project. Additionally, we observed differences between training and testing efforts; while training samples were more equally distributed for Team Makeup, the distribution was less equal for their testing data, with Glossier and Neutrogena accounting for a smaller amount of their overall testing data (9.2\% and 5\% respectively). In this section, we describe why these differences emerged as a result of decisions learners made about which labels to add data to.

An example of shifting away from equally distributed training data was observed with Team Makeup. While the team originally decided to add 100 images for each label, they soon realized this was a sub-optimal strategy because the model had more difficulty identifying some labels versus others. In response, their team only add data to labels their model did not work as well for:

\begin{quote}
We first decided to have a set number of photos to add for each label – 100 for each label just as a base point. But then afterward, we realized that that wasn't the best way to go about it because [for] some objects, the model had more difficulties identifying than others. And so we ended up just taking more photos for the ones that were difficult to classify.    
\end{quote}

The strategy to add mostly to labels with lower classification accuracy was also taken on by Teams Plants and Fashion, who both described this strategy as attending to ``equity over equality.''  Participant Plants-P-1 shared that, ``My team really value[s] the mindset of equity over equality when it came to training labels. So once we realize that one label was working most of the time, or even working 100\% of the time, we wanted to focus on the labels that maybe weren't working as much and that really helped us.''  They determined they needed more data for leafy green plants compared to their only flower label, a white and purple orchid plant; their orchid label represented 8.7\% of their total training data, compared to their Pothos label which had 29.7\%. Because the orchid plant was visually distinct from the other labels, and the model was consistently able to classify it correctly, the team members turned their attention to labels that were not performing as well, as described by Plants-P2: “Our orchid label had a lot less data than anything else because it was so different than all the other plants, and so the model was able to recognize it with a lot less data than the other plants that were all kind of green and similar shapes.” Likewise, Team Fashion described how more data was needed for brand logos that looked visually similar to one another compared to those that were unique and easier for the model to identify.  

Thus, by attending to model performance, participants wrestled with the machine learnability of individual classes, ultimately prioritizing adding more data to underperforming labels over having more evenly distributed class balance.  

\subsection{DDP4: Inspecting for data quality}
\label{sec:results-ddp4}

Participants discovered and interrogated data quality issues by reasoning about misclassifications that surfaced in the testing interface of \system. Misclassifications were largely project-specific, stemming from unique qualities of the objects used in their data.  To illustrate how collaboration played a role in how teams resolved data quality issues, we provide an in-depth description of a debugging scenario faced by Team Fashion, followed by a comparison of how this group's approach was reflected across other teams.

\subsubsection{Vignette: Resolving Data Quality Issues}
\label{result:ff-case-study}

Team Fashion designed an application to help consumers learn about the sustainability of different clothing brands, creating a model that could identify five common brands and display a sustainability rating from 1-5 (using data from Good On You \cite{goodOnYou}), along with suggestions for second-hand shopping or more sustainable alternatives.  Their data incorporated photos of physical garments they owned as well as images from the Internet of different colored clothing.

\begin{figure}[h]
  \centering
  \includegraphics[width=0.75\linewidth]{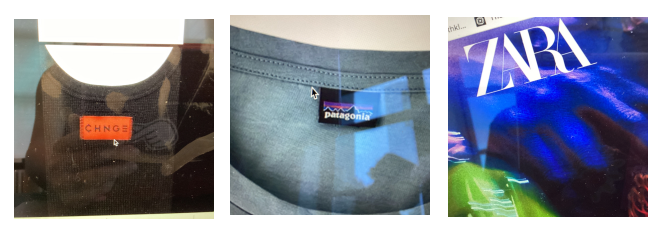}
  \caption{Examples of removed training images from FF’s dataset that included reflections from a laptop screen}
  \Description{}
  \label{fig:ff-reflections}
\end{figure}

Initially, members of Team Fashion captured images where each article of clothing was fully visible, but they quickly transitioned to taking more close-up shots of the brand logos themselves. This change in strategy came about because of the model’s low accuracy when testing, leading the team to look for issues in their dataset using the training and testing dashboards in \system.  Inspecting their test data led them to discover a pattern of reflective artifacts in images of Internet search results they photographed from a laptop screen (Figure \ref{fig:ff-reflections}). This problem was first brought up by participant Fashion-P2 based on a testing result – in response, the team examined the training data for similar patterns, hypothesizing that, “Maybe it [the model] was associating whatever has reflections with that brand that had that [reflections] in the training data.”  After identifying and discussing the issue together, they then they then removed approximately 20 images from their training data, across three of their labels as shown in Figure \ref{fig:ff-reflections}.

In our post-camp debriefs, Fashion-P1 described their team’s debugging process as follows:

\begin{quote}
    I think we each kind of gave different insights. We all kind of contributed to it. We were just looking at it [the training data] on our individual iPad's. But then as we kept studying the failure cases, I think each of us came up with, like, “Oh, this could be a reflection,” or, “Oh, this could be like this color”... We came up with a couple more theories...[and] the two main ones that we all kind of came to a conclusion on [were] after seeing our individual testing data and confirming that...this seems to be like a pattern in all of them.
\end{quote}

Notably, it was through a combination of reviewing their training and testing data in \system along with group discussion about patterns in misclassified images that the team decided to clean their dataset.  Describing this process during their team's final presentation, Fashion-C shared: ``We spent a lot of time analyzing our failure cases and looking for problematic spurious relationships. This allowed us to identify which pieces of data had to be removed and which ones had to be increased.''

Analyzing failed test samples as a group also provided space for each team member’s perspective to be valued, as shared by Fashion-P1 ``I feel like all of us just approached the situation in different ways, and we thought of creative ways that we could test our models. So it's definitely so much better than working alone because you can bring in so many new perspectives and kind of be more efficient overall.''  Through discussing the model's data, the team was able to share multiple theories about qualities that may contribute to misclassification, and ultimately act on those theories by removing lower quality images from their dataset.

\subsubsection{Data Quality Issues Across Projects}

While Team Fashion's data cleaning process resulted from collaborative discussion, \system also supported asynchronous data cleaning by supporting ambient awareness of the dataset. Describing this process, Nutrition-P1 shared, “I think we would look at each other’s to see if they were like 100\% correct. And if the data was accurate – to make sure there weren’t any really blurry photos…or maybe the wrong item...I feel like [we] just looked over each other’s work just to make sure it was correct.”  Team Makeup echoed this idea, with participant Makeup-P2 describing a scenario where she had accidentally collected images under the wrong label, and her teammates helped her delete the mislabeled data, stating that this saved her group time and helped them work efficiently. These examples show how debugging related to data quality was facilitated by having multiple eyes on the data.

These scenarios point to how a multi-user modeling experience with \system enabled learners to monitor data quality by reviewing each other's images, discussing and debating theories about what might lead to model failures, and collectively sharing in the responsibility to delete lower quality images. Throughout this process, the testing and training dashboards in \system anchored their discussions as they continually revised their datasets.

\subsection{Increased Confidence in Dataset Design}

Throughout the camp, we asked participants daily to rate their ability to contribute to conversations about ``how data can influence model performance'', ``the design and improvement of ML systems'', and ``examples of ML to problem solve''. These ratings were  on a scale from 1 (strongly disagree) to 5 (strongly agree), and we compared the mean of each participant's response to these questions at the beginning and end of the camp. This showed an increase across the camp of about half a point (0.53), from 3.9 (\textit{SD}=.67) on the first day to 4.8 (\textit{SD}=.43) on the final day. This increase was statistically significant, \textit{t}(18) = 5.48, \textit{p} < .001. The effect size of this difference, using Cohen's \textit{d}, was 1.26, which is considered a ``large'' effect \cite{cohen1988statistical}. This shows that participants' confidence in dataset design had increased, because understanding how data can influence model performance and the design and improvement of ML systems are both fundamental to dataset design.

Participants reflected that their ability to modify datasets directly, and see how those changes affected model performance, was made possible through their hands-on experience using \system. Participant Fashion-P1 expressed this idea by sharing, ``Working with an image classifier was really cool in that we could modify our objects any way that we wanted in order to train the model and make it more accurate.'' Similarly, participant Foodwaste-P3 described how \system supported her experimentation with data: ``When we had an issue, we were able to look at our training data and [see if] `It didn't work against this background because we didn't have it in our training data.' The issues we did have, there were solutions to it --- just add more training data or delete some data.'' In turn, participants were able to take on an active role in building models, which student Foodwaste-P2 contrasted with only being on the receiving end of ML systems: ``It was really cool to be able to be on the developer's side to train and test models that we had only been users for before.''
\section{DISCUSSION}

Our analysis of group ML model building using \system, situated in the designed context of the \campname, revealed how collaboration positively shaped the enactment of core dataset design practices. These results demonstrate that the ways participants explored the design of ML datasets during the camp was compatible with our initial design goals for the \system app.

\subsection{The Role of Collaboration in Designing and Debugging Data}

Collaboration influenced how participants engaged in each of the dataset design practices, playing a key role throughout their model building process by helping participants 1) collectively inspect data to identify differences between misclassified and correctly classified test data; 2) discuss ideas about the cause of misclassifications; and 3) coordinate how to enact those ideas by modifying data and retraining their models together. 

We observed that, through continuous discourse, individuals encountered ways to capture data other than what they had individually considered, which helped them create more diverse training data sets (DDP1). We saw individual differences in data collection arise for Teams Foodwaste and Nutrition, who both discovered that team members differed in whether they photographed their objects with hands visible. Discovery of these differences emerged through both active verbal discussion and ambient progress monitoring of their shared training and testing dashboards. By noticing and discussing the ``hands versus no hands'' conditions, participants were able to propose, test, and revise ideas about how and to what extent human hands impacted their model's performance (DDP2). 

Conversations between participants also deepened their understanding of data quality and its relationship to model performance, similar to prior findings that peer-to-peer dialogue can support students in generating more robust, diverse ideas when they use ML tools \cite{kaspersen2021votestratesml} and that acknowledging and discussing proposed ideas can lead to positive uptake of new solutions \cite{barron2003smart}. As exemplified by members of Team Fashion (as described in \ref{sec:results-ddp4}), \system surfaced discussion about the root causes of misclassified test images. By working together, participants brainstormed reasons why certain images might be misclassified and the specific data quality issues that may have contributed to erroneous predictions. They concluded that unintentional image artifacts led to poor data quality (DDP4), deleted these images from their dataset, and retrained their model to improve its performance (DDP2). \system's data editing features supported their revisions by enabling data deletion, an atypical practice for novice ML-users \cite{yang2018grounding}. 

Our results show how enabling collaboration in a novice ML modeling experience positively contribute to learning about ML and data diversity, quality, and model performance.

\subsection{The Value of Testing Interfaces for Encouraging Model Iteration}

All teams iteratively refined their models, retraining them on average 40 times. Model iteration was supported by a key feature of the \system experience that is missing in other novice-oriented ML tools: a testing interface for collecting and reviewing test datasets. Existing tools (as described in \ref{sec:ml-education-k-12}) largely rely on ephemeral live classification, where users monitor model results in real-time, providing little opportunity for users to revisit individual misclassified samples, compare misclassifications with correctly classified data, and determine whether evaluation results improve in response to model retraining. In contrast, we observed that the testing interface in \system was commonly used by students to monitor their model's progress, encouraging them to revisit and revise their datasets, an iterative practice common among ML professionals but overlooked by ML novices \cite{yang2018grounding}.

Because participants iteratively refined their datasets and retrained their models, they needed to manage multiple dataset considerations in parallel, just as ML practitioners do when dealing with real-world, messy data \cite{sambasivan2021everyone}. For example, when groups attended to underperforming labels, they described how models may not necessarily need the same amount of data per label (one strategy for class balance). Specifically, they reasoned that more data might be needed for labels that are harder for a model to learn, referring to the learnability of a label. In this way, participants were authentically engaging with dataset design practices because they could quickly update their data and test what impact changes to their data had on model performance (DDP2). As described by participant Fashion-P1,

\begin{quote}
It can be really hard to build a machine learning model because of many factors that play into it... But using \system allowed us to really accurately visualize what that would actually look like, because we were able kind of see how bad data could affect the model and how to remove that.
\end{quote} 

With \system, we show how testing interfaces supported iterative data refinement and model retraining. Because model iteration is such a foundational part of ML model design \cite{hohman2020understanding}, we suggest that other ML tools and educational efforts would benefit from incorporating user experiences around refining and reviewing test data to encourage more model iteration.

\subsection{Learner-driven Projects for Authentic ML Modeling}

Because participants chose their own projects and used their own data, each group encountered unique considerations around model use that shaped their thinking about what representative data means. Iteratively refining their datasets and assessing how that affected model performance (DDP 2), alongside the use of personally-designed data during the ML camp, enabled participants to explore aspects of dataset diversity (DDP1) that were not possible in prior work because learners either used the same underlying materials \cite{dwivedi2021exploring} or their project ideas were  selected for them based on feasibility \cite{vartiainen2020machine}. 

While previous work \cite{dwivedi2021exploring} described learners capturing data from multiple backgrounds and angles (the \textit{perspectives} and \textit{contexts} described in our dataset diversity strategies in Section \ref{sec:results-ddp2}), we saw two additional strategies around \textit{states} and \textit{types}, where participants considered multiple form factors or conditions for a given object, as well as the ways a class of objects might be represented. We believe that the use of personal data may provide opportunities for encountering domain-specific edge cases for model use, such as the use of healthy and sick plants for the Team Plants. This can enrich learner understanding and appreciation for how datasets should accurately reflect a variety of user scenarios.

Further, because each team applied ML to distinct problems, participants were able to expand their notion of how ML can applied to range of issues in the world. One participant described this by stating, ``I love how we can utilize \system and AI and ML in general across a variety and spectrum of topics.'' The cultural relevancy of the projects and data that teams designed is especially important given prior work suggesting that engaging in socially-relevant projects can increase participation of female-identifying students in CS \cite{khan2016computing}. \campname participants not only built personally-relevant models, but embedded these models into custom apps, allowing them to realize how models can be integrated into user experiences. This end-to-end model and application design experience helped participants not only develop an understanding of data in ML, but also see their own role in shaping ML systems in the future. Reflecting on her experience, participant Foodwaste-P2 shared,

\begin{quote}
In the lessons before we started using \system, we heard about everyday uses of AI and different models like, Instagram [and] Snapchat filters. I thought it was really cool to be able to use our own objects and then actually train and test and go through all of that stuff that developers have to go through to create the things that we use.
\end{quote}

Through using \system in the ML camps, learners gained confidence and saw themselves as developers of future ML technologies with social impact. This was demonstrated by responses to the daily surveys, which showed a statistically significant increase in confidence in contributing to conversations about the design and improvement of ML systems with a large effect size from the first to the last day of the camp.

\section{Limitations and Future Work}
\label{sec:limitations-future-work}

The use of \system we describe in this paper was necessarily shaped by the \campname context – an out-of-school enrichment experience designed primarily to support young women and gender nonconforming youth. The \campname centered activities and discussions at the intersection of technology and social equity, and this lens likely played a role in participants picking socially-minded final projects. Further, as a supportive environment for youth underrepresented in technology, \campname may have especially fostered a safe place for participants to explore topics of personal relevance such as sustainable fashion and racial inclusion in the beauty industry. While these topics highlight the importance of creating personally relevant applications, in this paper we did not investigate student identity development in relation to gender and computing. Given the lack of representation in the technology industry today, we believe further research on \campname is needed to better understand how the intentional design of a learning environment for female and gender nonconforming youth supported students' engagement with \system. We also acknowledge that future work is required to determine the transferability of our results to other contexts such as in-school learning experiences and learners of mixed gender and cultural identities.

As a high percentage (78\%) of the camp participants were alumni of other \nonprofit summer camps, participants were likely to have more experience with computing than the average \nonprofit student, as well as more experience working in groups to develop projects (as all \nonprofit camps involve students working in groups to develop final projects). We note that participants in our study did have a higher rate of free or reduced lunch (65\% for study participants compared to 50\% for \nonprofit participants in all camps), and 93\% of the study participants identified as people of color (compared to 82\% for \nonprofit participants in all camps).

The majority of images that participants captured included items they could reasonably bring to the camp themselves, as they were unable to take their iPads home and thus use \system outside of the camp. While we did see participants diversify their datasets to align with their imagined user scenarios (such as capturing images from the web to expand their dataset), we imagine that leveraging the mobile quality of iPads to collect data in multiple different contexts, in and outside a classroom, may further learning and experimentation with data diversity.

While decisions made by students related to data diversity (DDP1), model evaluation (DDP2), and data quality (DDP4) have ethical implications \cite{mitchell2020diversity}, we acknowledge that our analyses did not focus on these. The ethical and critical issues that students encounter when using \system  warrant a comprehensive exploration and analysis. Future research should explore how students' ideas of justice and ethical stances influence their decisions when engaging with data design practices, and how instructional activities may support critical engagement with \system \cite{ali2019constructionism, irgens2022characterizing}. 

Our research team was only able to closely shadow three of the six final project teams. As a result, for the other three teams, we were reliant on participants' reflections on their experience, either through design journals they maintained during camp or post-camp interviews. To the best of our abilities, we tried to corroborate the issues they self-described using our log visualizations, but for teams that were not directly observed, there may have been debugging moments that were not recollected in participant reflections and thus not represented in our analyses.

The \campname camp was our first test of our in-app logging system, and we ran into technical difficulties that precluded our logging of when users deleted individual images in their datasets. Missing logs of image deletions gave us less insight into how individuals may have taken on data cleaning responsibilities throughout final projects. While our team captured supplemental snapshots of students' Co-ML projects and corresponding training and testing data twice a day when teams were building their final projects, this information only provides context about deleted images at the group-level and at specific moments in time; we are unable to attribute deletion of specific images to particular individuals moment-to-moment. This makes it difficult for us to definitively identify what information in the app (such as a misclassified test sample) might have motivated users to delete images in their dataset. Further, we had limited capability to analyze qualitative factors of deleted images because we do not retain those images to protect user privacy. We mention these challenges because handling of deleted participant-collected data is especially pertinent to ML education, where considerations about data ownership, privacy, and ethics are important. Screen recordings or videos of participant interactions during the model-building process may be especially valuable to understanding how learners' collaboratively cleaned data, neither of which our team was able to collect in this pilot study.

Finally, we see an opportunity in future work with \system to analyze and categorize different types of collaboration and its impact on learning of ML. Our initial analysis of log data revealed two basic collaborative strategies for data collection. Some teams took a "divide and conquer" approach, where data collection responsibilities were split between team members, with members focused solely on a subset of labels. Other teams had a "cross-label" approach where group members had contributed to most labels by the end of the project. In the design of \system, we envisioned that reviewing and contributing across most or all labels of a project would provide more opportunities for participants to discuss data that others had added, and to realize different perspectives for diversifying their dataset. However, further work is needed to fully relate our log data with participant and group motivations for working more narrowly or expansively across labels; we believe this future work may be able to identify specific collaborative behaviors and strategies that support learning about the relationship between data and model performance.
\section{ACKNOWLEDGMENTS}
We would like to thank all \campname participants who enrolled in our research study for sharing their time and feedback with us. Many members of the \nonprofitfull team played an instrumental role in organizing and implementing the \campname, including Tara Tran, Laura Angelich, Hallie Smith, Dorothy Chang, Hannah Kim, and Allie Feldman. At Apple, we would like to thank Stuart Ralston, Richard Lombardo, Casandra Sisneros, Mike Mead, Adriana Hilliard, Paris Garrett, and Emmanuel Adepoju for all of their support in facilitating the \nonprofitfull partnership, educator training, engineering of \system, and student support through the camps.

\bibliographystyle{ACM-Reference-Format}
\bibliography{bibliography.bib}


\begin{thebibliography}{85}


\ifx \showCODEN    \undefined \def \showCODEN     #1{\unskip}     \fi
\ifx \showDOI      \undefined \def \showDOI       #1{#1}\fi
\ifx \showISBNx    \undefined \def \showISBNx     #1{\unskip}     \fi
\ifx \showISBNxiii \undefined \def \showISBNxiii  #1{\unskip}     \fi
\ifx \showISSN     \undefined \def \showISSN      #1{\unskip}     \fi
\ifx \showLCCN     \undefined \def \showLCCN      #1{\unskip}     \fi
\ifx \shownote     \undefined \def \shownote      #1{#1}          \fi
\ifx \showarticletitle \undefined \def \showarticletitle #1{#1}   \fi
\ifx \showURL      \undefined \def \showURL       {\relax}        \fi
\providecommand\bibfield[2]{#2}
\providecommand\bibinfo[2]{#2}
\providecommand\natexlab[1]{#1}
\providecommand\showeprint[2][]{arXiv:#2}

\bibitem[\protect\citeauthoryear{Agassi, Erel, Wald, and Zuckerman}{Agassi et~al\mbox{.}}{2019}]%
        {agassi2019scratch}
\bibfield{author}{\bibinfo{person}{Adam Agassi}, \bibinfo{person}{Hadas Erel}, \bibinfo{person}{Iddo~Yehoshua Wald}, {and} \bibinfo{person}{Oren Zuckerman}.} \bibinfo{year}{2019}\natexlab{}.
\newblock \showarticletitle{Scratch nodes ML: A playful system for children to create gesture recognition classifiers}. In \bibinfo{booktitle}{\emph{Extended Abstracts of the 2019 CHI Conference on Human Factors in Computing Systems}}. \bibinfo{pages}{1--6}.
\newblock


\bibitem[\protect\citeauthoryear{Ali, Payne, Williams, Park, and Breazeal}{Ali et~al\mbox{.}}{2019}]%
        {ali2019constructionism}
\bibfield{author}{\bibinfo{person}{Safinah Ali}, \bibinfo{person}{Blakeley~H Payne}, \bibinfo{person}{Randi Williams}, \bibinfo{person}{Hae~Won Park}, {and} \bibinfo{person}{Cynthia Breazeal}.} \bibinfo{year}{2019}\natexlab{}.
\newblock \showarticletitle{Constructionism, ethics, and creativity: Developing primary and middle school artificial intelligence education}. In \bibinfo{booktitle}{\emph{International workshop on education in artificial intelligence k-12 (eduai’19)}}, Vol.~\bibinfo{volume}{2}. \bibinfo{pages}{1--4}.
\newblock


\bibitem[\protect\citeauthoryear{Apple}{Apple}{2022}]%
        {createml}
\bibfield{author}{\bibinfo{person}{Apple}.} \bibinfo{year}{2022}\natexlab{}.
\newblock \bibinfo{title}{CreateML}.
\newblock
\newblock
\urldef\tempurl%
\url{https://developer.apple.com/machine-learning/create-ml/}
\showURL{%
\tempurl}


\bibitem[\protect\citeauthoryear{Arastoopour~Irgens, Adisa, Bailey, and Quesada}{Arastoopour~Irgens et~al\mbox{.}}{2022a}]%
        {irgens2022designing}
\bibfield{author}{\bibinfo{person}{Golnaz Arastoopour~Irgens}, \bibinfo{person}{Ibrahim Adisa}, \bibinfo{person}{Cinamon Bailey}, {and} \bibinfo{person}{Hazel~Vega Quesada}.} \bibinfo{year}{2022}\natexlab{a}.
\newblock \showarticletitle{Designing with and for Youth: A Participatory Design Research Approach for Critical Machine Learning Education}.
\newblock \bibinfo{journal}{\emph{Educational Technology \& Society}} \bibinfo{volume}{25}, \bibinfo{number}{4} (\bibinfo{year}{2022}), \bibinfo{pages}{126--141}.
\newblock


\bibitem[\protect\citeauthoryear{Arastoopour~Irgens, Vega, Adisa, and Bailey}{Arastoopour~Irgens et~al\mbox{.}}{2022b}]%
        {irgens2022characterizing}
\bibfield{author}{\bibinfo{person}{Golnaz Arastoopour~Irgens}, \bibinfo{person}{Hazel Vega}, \bibinfo{person}{Ibrahim Adisa}, {and} \bibinfo{person}{Cinamon Bailey}.} \bibinfo{year}{2022}\natexlab{b}.
\newblock \showarticletitle{Characterizing children’s conceptual knowledge and computational practices in a critical machine learning educational program}.
\newblock \bibinfo{journal}{\emph{International Journal of Child-Computer Interaction}}  \bibinfo{volume}{34} (\bibinfo{year}{2022}), \bibinfo{pages}{100541}.
\newblock


\bibitem[\protect\citeauthoryear{Babbage}{Babbage}{2022}]%
        {babbage2022passages}
\bibfield{author}{\bibinfo{person}{Charles Babbage}.} \bibinfo{year}{2022}\natexlab{}.
\newblock \bibinfo{booktitle}{\emph{Passages from the Life of a Philosopher}}.
\newblock \bibinfo{publisher}{DigiCat}.
\newblock


\bibitem[\protect\citeauthoryear{Barron}{Barron}{2003}]%
        {barron2003smart}
\bibfield{author}{\bibinfo{person}{Brigid Barron}.} \bibinfo{year}{2003}\natexlab{}.
\newblock \showarticletitle{When smart groups fail}.
\newblock \bibinfo{journal}{\emph{The journal of the learning sciences}} \bibinfo{volume}{12}, \bibinfo{number}{3} (\bibinfo{year}{2003}), \bibinfo{pages}{307--359}.
\newblock


\bibitem[\protect\citeauthoryear{Bigman, Roy, Garcia, Suzara, Wang, and Piech}{Bigman et~al\mbox{.}}{2021}]%
        {bigman2021pearprogram}
\bibfield{author}{\bibinfo{person}{Maxwell Bigman}, \bibinfo{person}{Ethan Roy}, \bibinfo{person}{Jorge Garcia}, \bibinfo{person}{Miroslav Suzara}, \bibinfo{person}{Kaili Wang}, {and} \bibinfo{person}{Chris Piech}.} \bibinfo{year}{2021}\natexlab{}.
\newblock \showarticletitle{PearProgram: A more fruitful approach to pair programming}. In \bibinfo{booktitle}{\emph{Proceedings of the 52nd ACM Technical Symposium on Computer Science Education}}. \bibinfo{pages}{900--906}.
\newblock


\bibitem[\protect\citeauthoryear{Blumenfeld, Marx, Soloway, and Krajcik}{Blumenfeld et~al\mbox{.}}{1996}]%
        {blumenfeld1996learning}
\bibfield{author}{\bibinfo{person}{Phyllis~C Blumenfeld}, \bibinfo{person}{Ronald~W Marx}, \bibinfo{person}{Elliot Soloway}, {and} \bibinfo{person}{Joseph Krajcik}.} \bibinfo{year}{1996}\natexlab{}.
\newblock \showarticletitle{Learning with peers: From small group cooperation to collaborative communities}.
\newblock \bibinfo{journal}{\emph{Educational researcher}} \bibinfo{volume}{25}, \bibinfo{number}{8} (\bibinfo{year}{1996}), \bibinfo{pages}{37--39}.
\newblock


\bibitem[\protect\citeauthoryear{Bransford, Brown, Cocking, et~al\mbox{.}}{Bransford et~al\mbox{.}}{2000}]%
        {bransford2000people}
\bibfield{author}{\bibinfo{person}{John~D Bransford}, \bibinfo{person}{Ann~L Brown}, \bibinfo{person}{Rodney~R Cocking}, {et~al\mbox{.}}} \bibinfo{year}{2000}\natexlab{}.
\newblock \bibinfo{booktitle}{\emph{How people learn}}. Vol.~\bibinfo{volume}{11}.
\newblock \bibinfo{publisher}{Washington, DC: National academy press}.
\newblock


\bibitem[\protect\citeauthoryear{Braught, Wahls, and Eby}{Braught et~al\mbox{.}}{2011}]%
        {braught2011case}
\bibfield{author}{\bibinfo{person}{Grant Braught}, \bibinfo{person}{Tim Wahls}, {and} \bibinfo{person}{L~Marlin Eby}.} \bibinfo{year}{2011}\natexlab{}.
\newblock \showarticletitle{The case for pair programming in the computer science classroom}.
\newblock \bibinfo{journal}{\emph{ACM Transactions on Computing Education (TOCE)}} \bibinfo{volume}{11}, \bibinfo{number}{1} (\bibinfo{year}{2011}), \bibinfo{pages}{1--21}.
\newblock


\bibitem[\protect\citeauthoryear{Buolamwini and Gebru}{Buolamwini and Gebru}{2018}]%
        {buolamwini2018gender}
\bibfield{author}{\bibinfo{person}{Joy Buolamwini} {and} \bibinfo{person}{Timnit Gebru}.} \bibinfo{year}{2018}\natexlab{}.
\newblock \showarticletitle{Gender shades: Intersectional accuracy disparities in commercial gender classification}. In \bibinfo{booktitle}{\emph{Conference on fairness, accountability and transparency}}. PMLR, \bibinfo{pages}{77--91}.
\newblock


\bibitem[\protect\citeauthoryear{Carney, Webster, Alvarado, Phillips, Howell, Griffith, Jongejan, Pitaru, and Chen}{Carney et~al\mbox{.}}{2020}]%
        {carney2020teachable}
\bibfield{author}{\bibinfo{person}{Michelle Carney}, \bibinfo{person}{Barron Webster}, \bibinfo{person}{Irene Alvarado}, \bibinfo{person}{Kyle Phillips}, \bibinfo{person}{Noura Howell}, \bibinfo{person}{Jordan Griffith}, \bibinfo{person}{Jonas Jongejan}, \bibinfo{person}{Amit Pitaru}, {and} \bibinfo{person}{Alexander Chen}.} \bibinfo{year}{2020}\natexlab{}.
\newblock \showarticletitle{Teachable machine: Approachable Web-based tool for exploring machine learning classification}. In \bibinfo{booktitle}{\emph{Extended Abstracts of the 2020 CHI Conference on Human Factors in Computing Systems}}. \bibinfo{pages}{1--8}.
\newblock


\bibitem[\protect\citeauthoryear{Charmaz}{Charmaz}{2000}]%
        {charmaz2000grounded}
\bibfield{author}{\bibinfo{person}{Kathy Charmaz}.} \bibinfo{year}{2000}\natexlab{}.
\newblock \showarticletitle{Grounded theory: Objectivist and constructivist methods}.
\newblock \bibinfo{journal}{\emph{Handbook of qualitative research}} \bibinfo{volume}{2}, \bibinfo{number}{1} (\bibinfo{year}{2000}), \bibinfo{pages}{509--535}.
\newblock


\bibitem[\protect\citeauthoryear{Cohen}{Cohen}{1988}]%
        {cohen1988statistical}
\bibfield{author}{\bibinfo{person}{Jacob Cohen}.} \bibinfo{year}{1988}\natexlab{}.
\newblock \bibinfo{booktitle}{\emph{Statistical power analysis for the behavioral sciences}}.
\newblock \bibinfo{publisher}{Routledge}.
\newblock


\bibitem[\protect\citeauthoryear{Dastin}{Dastin}{2018}]%
        {dastin2018amazon}
\bibfield{author}{\bibinfo{person}{Jeffrey Dastin}.} \bibinfo{year}{2018}\natexlab{}.
\newblock \showarticletitle{Amazon scraps secret AI recruiting tool that showed bias against women}.
\newblock In \bibinfo{booktitle}{\emph{Ethics of data and analytics}}. \bibinfo{publisher}{Auerbach Publications}, \bibinfo{pages}{296--299}.
\newblock


\bibitem[\protect\citeauthoryear{Deitrick, O'Connell, and Shapiro}{Deitrick et~al\mbox{.}}{2014}]%
        {deitrick2014discourse}
\bibfield{author}{\bibinfo{person}{Elise Deitrick}, \bibinfo{person}{Brian O'Connell}, {and} \bibinfo{person}{R~Benjamin Shapiro}.} \bibinfo{year}{2014}\natexlab{}.
\newblock \showarticletitle{The discourse of creative problem solving in childhood engineering education}.
\newblock \bibinfo{publisher}{Boulder, CO: International Society of the Learning Sciences}.
\newblock


\bibitem[\protect\citeauthoryear{Deitrick, Shapiro, and Gravel}{Deitrick et~al\mbox{.}}{2016}]%
        {deitrick2016we}
\bibfield{author}{\bibinfo{person}{Elise Deitrick}, \bibinfo{person}{R~Benjamin Shapiro}, {and} \bibinfo{person}{Brian Gravel}.} \bibinfo{year}{2016}\natexlab{}.
\newblock \showarticletitle{How do we assess equity in programming pairs?}
\newblock \bibinfo{publisher}{Singapore: International Society of the Learning Sciences}.
\newblock


\bibitem[\protect\citeauthoryear{Dourish and Bellotti}{Dourish and Bellotti}{1992}]%
        {dourish1992awareness}
\bibfield{author}{\bibinfo{person}{Paul Dourish} {and} \bibinfo{person}{Victoria Bellotti}.} \bibinfo{year}{1992}\natexlab{}.
\newblock \showarticletitle{Awareness and coordination in shared workspaces}. In \bibinfo{booktitle}{\emph{Proceedings of the 1992 ACM conference on Computer-supported cooperative work}}. \bibinfo{pages}{107--114}.
\newblock


\bibitem[\protect\citeauthoryear{Druga}{Druga}{2018}]%
        {druga2018growing}
\bibfield{author}{\bibinfo{person}{Stefania Druga}.} \bibinfo{year}{2018}\natexlab{}.
\newblock \emph{\bibinfo{title}{Growing up with AI: Cognimates: from coding to teaching machines}}.
\newblock \bibinfo{thesistype}{Ph.D. Dissertation}. \bibinfo{school}{Massachusetts Institute of Technology}.
\newblock


\bibitem[\protect\citeauthoryear{Druga, Christoph, and Ko}{Druga et~al\mbox{.}}{2022a}]%
        {druga2022family}
\bibfield{author}{\bibinfo{person}{Stefania Druga}, \bibinfo{person}{Fee~Lia Christoph}, {and} \bibinfo{person}{Amy~J Ko}.} \bibinfo{year}{2022}\natexlab{a}.
\newblock \showarticletitle{Family as a Third Space for AI Literacies: How do children and parents learn about AI together?}. In \bibinfo{booktitle}{\emph{Proceedings of the 2022 CHI Conference on Human Factors in Computing Systems}}. \bibinfo{pages}{1--17}.
\newblock


\bibitem[\protect\citeauthoryear{Druga, Otero, and Ko}{Druga et~al\mbox{.}}{2022b}]%
        {druga2022landscape}
\bibfield{author}{\bibinfo{person}{Stefania Druga}, \bibinfo{person}{Nancy Otero}, {and} \bibinfo{person}{Amy~J Ko}.} \bibinfo{year}{2022}\natexlab{b}.
\newblock \showarticletitle{The landscape of teaching resources for ai education}. In \bibinfo{booktitle}{\emph{Proceedings of the 27th ACM Conference on on Innovation and Technology in Computer Science Education Vol. 1}}. \bibinfo{pages}{96--102}.
\newblock


\bibitem[\protect\citeauthoryear{Dwivedi, Gandhi, Parikh, Coenraad, Bonsignore, and Kacorri}{Dwivedi et~al\mbox{.}}{2021}]%
        {dwivedi2021exploring}
\bibfield{author}{\bibinfo{person}{Utkarsh Dwivedi}, \bibinfo{person}{Jaina Gandhi}, \bibinfo{person}{Raj Parikh}, \bibinfo{person}{Merijke Coenraad}, \bibinfo{person}{Elizabeth Bonsignore}, {and} \bibinfo{person}{Hernisa Kacorri}.} \bibinfo{year}{2021}\natexlab{}.
\newblock \showarticletitle{Exploring Machine Teaching with Children}. In \bibinfo{booktitle}{\emph{2021 IEEE Symposium on Visual Languages and Human-Centric Computing (VL/HCC)}}. IEEE, \bibinfo{pages}{1--11}.
\newblock


\bibitem[\protect\citeauthoryear{Estevez, Garate, and Gra{\~n}a}{Estevez et~al\mbox{.}}{2019}]%
        {estevez2019gentle}
\bibfield{author}{\bibinfo{person}{Julian Estevez}, \bibinfo{person}{Gorka Garate}, {and} \bibinfo{person}{Manuel Gra{\~n}a}.} \bibinfo{year}{2019}\natexlab{}.
\newblock \showarticletitle{Gentle introduction to artificial intelligence for high-school students using scratch}.
\newblock \bibinfo{journal}{\emph{IEEE access}}  \bibinfo{volume}{7} (\bibinfo{year}{2019}), \bibinfo{pages}{179027--179036}.
\newblock


\bibitem[\protect\citeauthoryear{Eubanks}{Eubanks}{2018}]%
        {eubanks2018automating}
\bibfield{author}{\bibinfo{person}{Virginia Eubanks}.} \bibinfo{year}{2018}\natexlab{}.
\newblock \bibinfo{booktitle}{\emph{Automating inequality: How high-tech tools profile, police, and punish the poor}}.
\newblock \bibinfo{publisher}{St. Martin's Press}.
\newblock


\bibitem[\protect\citeauthoryear{Fiebrink}{Fiebrink}{2019}]%
        {fiebrink2019machine}
\bibfield{author}{\bibinfo{person}{Rebecca Fiebrink}.} \bibinfo{year}{2019}\natexlab{}.
\newblock \showarticletitle{Machine learning education for artists, musicians, and other creative practitioners}.
\newblock \bibinfo{journal}{\emph{ACM Transactions on Computing Education (TOCE)}} \bibinfo{volume}{19}, \bibinfo{number}{4} (\bibinfo{year}{2019}), \bibinfo{pages}{1--32}.
\newblock


\bibitem[\protect\citeauthoryear{for Democracy and Technology}{for Democracy and Technology}{2019}]%
        {cdt}
\bibfield{author}{\bibinfo{person}{Center for Democracy} {and} \bibinfo{person}{Technology}.} \bibinfo{year}{2019}\natexlab{}.
\newblock \bibinfo{title}{AI \& Machine Learning}.
\newblock \bibinfo{howpublished}{\url{https://cdt.org/ai-machine-learning/}}.
\newblock


\bibitem[\protect\citeauthoryear{Glaser and Strauss}{Glaser and Strauss}{2017}]%
        {glaser2017discovery}
\bibfield{author}{\bibinfo{person}{Barney~G Glaser} {and} \bibinfo{person}{Anselm~L Strauss}.} \bibinfo{year}{2017}\natexlab{}.
\newblock \bibinfo{booktitle}{\emph{Discovery of grounded theory: Strategies for qualitative research}}.
\newblock \bibinfo{publisher}{Routledge}.
\newblock


\bibitem[\protect\citeauthoryear{Google}{Google}{2023}]%
        {googleColab}
\bibfield{author}{\bibinfo{person}{Google}.} \bibinfo{year}{2023}\natexlab{}.
\newblock \bibinfo{title}{Colab}.
\newblock \bibinfo{howpublished}{\url{https://colab.research.google.com}}.
\newblock


\bibitem[\protect\citeauthoryear{Group and Lab}{Group and Lab}{2023}]%
        {dailycurriculum}
\bibfield{author}{\bibinfo{person}{MIT Media Lab Personal~Robots Group} {and} \bibinfo{person}{MIT~STEP Lab}.} \bibinfo{year}{2023}\natexlab{}.
\newblock \bibinfo{title}{DAILy Curriculum for Middle School Students}.
\newblock \bibinfo{howpublished}{\url{https://raise.mit.edu/daily/index.html}}.
\newblock


\bibitem[\protect\citeauthoryear{Hitron, Wald, Erel, and Zuckerman}{Hitron et~al\mbox{.}}{2018}]%
        {hitron2018introducing}
\bibfield{author}{\bibinfo{person}{Tom Hitron}, \bibinfo{person}{Iddo Wald}, \bibinfo{person}{Hadas Erel}, {and} \bibinfo{person}{Oren Zuckerman}.} \bibinfo{year}{2018}\natexlab{}.
\newblock \showarticletitle{Introducing children to machine learning concepts through hands-on experience}. In \bibinfo{booktitle}{\emph{Proceedings of the 17th ACM conference on interaction design and children}}. \bibinfo{pages}{563--568}.
\newblock


\bibitem[\protect\citeauthoryear{Hmelo-Silver, Jeong, Faulkner, and Hartley}{Hmelo-Silver et~al\mbox{.}}{2017}]%
        {hmelo2017computer}
\bibfield{author}{\bibinfo{person}{Cindy Hmelo-Silver}, \bibinfo{person}{Heisawn Jeong}, \bibinfo{person}{Roosevelt Faulkner}, {and} \bibinfo{person}{Kylie Hartley}.} \bibinfo{year}{2017}\natexlab{}.
\newblock \showarticletitle{Computer-supported collaborative learning in STEM domains: Towards a meta-synthesis}.
\newblock  (\bibinfo{year}{2017}).
\newblock


\bibitem[\protect\citeauthoryear{Hohman, Wongsuphasawat, Kery, and Patel}{Hohman et~al\mbox{.}}{2020}]%
        {hohman2020understanding}
\bibfield{author}{\bibinfo{person}{Fred Hohman}, \bibinfo{person}{Kanit Wongsuphasawat}, \bibinfo{person}{Mary~Beth Kery}, {and} \bibinfo{person}{Kayur Patel}.} \bibinfo{year}{2020}\natexlab{}.
\newblock \showarticletitle{Understanding and Visualizing Data Iteration in Machine Learning}. In \bibinfo{booktitle}{\emph{Proceedings of the SIGCHI Conference on Human Factors in Computing Systems}}. ACM.
\newblock
\urldef\tempurl%
\url{https://doi.org/10.1145/3313831.3376177}
\showDOI{\tempurl}


\bibitem[\protect\citeauthoryear{Hopkins, Hohman, Zappella, Cuadros, and Moritz}{Hopkins et~al\mbox{.}}{2023}]%
        {hopkins2023designing}
\bibfield{author}{\bibinfo{person}{Aspen Hopkins}, \bibinfo{person}{Fred Hohman}, \bibinfo{person}{Luca Zappella}, \bibinfo{person}{Xavier~Suau Cuadros}, {and} \bibinfo{person}{Dominik Moritz}.} \bibinfo{year}{2023}\natexlab{}.
\newblock \showarticletitle{Designing data: Proactive data collection and iteration for machine learning}.
\newblock \bibinfo{journal}{\emph{arXiv preprint arXiv:2301.10319}} (\bibinfo{year}{2023}).
\newblock


\bibitem[\protect\citeauthoryear{Jones and Fleming}{Jones and Fleming}{2013}]%
        {jones2013use}
\bibfield{author}{\bibinfo{person}{Danielle~L Jones} {and} \bibinfo{person}{Scott~D Fleming}.} \bibinfo{year}{2013}\natexlab{}.
\newblock \showarticletitle{What use is a backseat driver? A qualitative investigation of pair programming}. In \bibinfo{booktitle}{\emph{2013 IEEE Symposium on Visual Languages and Human Centric Computing}}. IEEE, \bibinfo{pages}{103--110}.
\newblock


\bibitem[\protect\citeauthoryear{Jordan, Devasia, Hong, Williams, and Breazeal}{Jordan et~al\mbox{.}}{2021}]%
        {jordan2021poseblocks}
\bibfield{author}{\bibinfo{person}{Brian Jordan}, \bibinfo{person}{Nisha Devasia}, \bibinfo{person}{Jenna Hong}, \bibinfo{person}{Randi Williams}, {and} \bibinfo{person}{Cynthia Breazeal}.} \bibinfo{year}{2021}\natexlab{}.
\newblock \showarticletitle{PoseBlocks: A toolkit for creating (and dancing) with AI}. In \bibinfo{booktitle}{\emph{Proceedings of the AAAI Conference on Artificial Intelligence}}, Vol.~\bibinfo{volume}{35}. \bibinfo{pages}{15551--15559}.
\newblock


\bibitem[\protect\citeauthoryear{Kahn and Winters}{Kahn and Winters}{2021}]%
        {kahn2021constructionism}
\bibfield{author}{\bibinfo{person}{Ken Kahn} {and} \bibinfo{person}{Niall Winters}.} \bibinfo{year}{2021}\natexlab{}.
\newblock \showarticletitle{Constructionism and AI: A history and possible futures}.
\newblock \bibinfo{journal}{\emph{British Journal of Educational Technology}} \bibinfo{volume}{52}, \bibinfo{number}{3} (\bibinfo{year}{2021}), \bibinfo{pages}{1130--1142}.
\newblock


\bibitem[\protect\citeauthoryear{Kaspersen, Bilstrup, Van~Mechelen, Hjorth, Bouvin, and Petersen}{Kaspersen et~al\mbox{.}}{2021}]%
        {kaspersen2021votestratesml}
\bibfield{author}{\bibinfo{person}{Magnus~Hoeholt Kaspersen}, \bibinfo{person}{Karl-Emil~Kjaer Bilstrup}, \bibinfo{person}{Maarten Van~Mechelen}, \bibinfo{person}{Arthur Hjorth}, \bibinfo{person}{Niels~Olof Bouvin}, {and} \bibinfo{person}{Marianne~Graves Petersen}.} \bibinfo{year}{2021}\natexlab{}.
\newblock \showarticletitle{VotestratesML: A high school learning tool for exploring machine learning and its societal implications}. In \bibinfo{booktitle}{\emph{FabLearn Europe/MakeEd 2021-An international conference on computing, design and making in education}}. \bibinfo{pages}{1--10}.
\newblock


\bibitem[\protect\citeauthoryear{Khan and Luxton-Reilly}{Khan and Luxton-Reilly}{2016}]%
        {khan2016computing}
\bibfield{author}{\bibinfo{person}{Nazish~Zaman Khan} {and} \bibinfo{person}{Andrew Luxton-Reilly}.} \bibinfo{year}{2016}\natexlab{}.
\newblock \showarticletitle{Is computing for social good the solution to closing the gender gap in computer science?}. In \bibinfo{booktitle}{\emph{Proceedings of the Australasian Computer Science Week Multiconference}}. \bibinfo{pages}{1--5}.
\newblock


\bibitem[\protect\citeauthoryear{Koesten, Gregory, Groth, and Simperl}{Koesten et~al\mbox{.}}{2021}]%
        {koesten2021talking}
\bibfield{author}{\bibinfo{person}{Laura Koesten}, \bibinfo{person}{Kathleen Gregory}, \bibinfo{person}{Paul Groth}, {and} \bibinfo{person}{Elena Simperl}.} \bibinfo{year}{2021}\natexlab{}.
\newblock \showarticletitle{Talking datasets--understanding data sensemaking behaviours}.
\newblock \bibinfo{journal}{\emph{International journal of human-computer studies}}  \bibinfo{volume}{146} (\bibinfo{year}{2021}), \bibinfo{pages}{102562}.
\newblock


\bibitem[\protect\citeauthoryear{Lane}{Lane}{2021}]%
        {lane2021machine}
\bibfield{author}{\bibinfo{person}{Dale Lane}.} \bibinfo{year}{2021}\natexlab{}.
\newblock \bibinfo{booktitle}{\emph{Machine learning for kids: A project-based introduction to artificial intelligence}}.
\newblock \bibinfo{publisher}{No Starch Press}.
\newblock


\bibitem[\protect\citeauthoryear{Lee, Gobir, Gurn, and Soep}{Lee et~al\mbox{.}}{2022}]%
        {lee2022black}
\bibfield{author}{\bibinfo{person}{Clifford~H Lee}, \bibinfo{person}{Nimah Gobir}, \bibinfo{person}{Alex Gurn}, {and} \bibinfo{person}{Elisabeth Soep}.} \bibinfo{year}{2022}\natexlab{}.
\newblock \showarticletitle{In the black mirror: Youth investigations into artificial intelligence}.
\newblock \bibinfo{journal}{\emph{ACM Transactions on Computing Education}} \bibinfo{volume}{22}, \bibinfo{number}{3} (\bibinfo{year}{2022}), \bibinfo{pages}{1--25}.
\newblock


\bibitem[\protect\citeauthoryear{Lee, Ali, Zhang, DiPaola, and Breazeal}{Lee et~al\mbox{.}}{2021}]%
        {lee2021developing}
\bibfield{author}{\bibinfo{person}{Irene Lee}, \bibinfo{person}{Safinah Ali}, \bibinfo{person}{Helen Zhang}, \bibinfo{person}{Daniella DiPaola}, {and} \bibinfo{person}{Cynthia Breazeal}.} \bibinfo{year}{2021}\natexlab{}.
\newblock \showarticletitle{Developing middle school students' AI literacy}. In \bibinfo{booktitle}{\emph{Proceedings of the 52nd ACM technical symposium on computer science education}}. \bibinfo{pages}{191--197}.
\newblock


\bibitem[\protect\citeauthoryear{Long and Magerko}{Long and Magerko}{2020}]%
        {long2020ai}
\bibfield{author}{\bibinfo{person}{Duri Long} {and} \bibinfo{person}{Brian Magerko}.} \bibinfo{year}{2020}\natexlab{}.
\newblock \showarticletitle{What is AI literacy? Competencies and design considerations}. In \bibinfo{booktitle}{\emph{Proceedings of the 2020 CHI Conference on Human Factors in Computing Systems}}. \bibinfo{pages}{1--16}.
\newblock


\bibitem[\protect\citeauthoryear{Long, Teachey, and Magerko}{Long et~al\mbox{.}}{2022}]%
        {long2022family}
\bibfield{author}{\bibinfo{person}{Duri Long}, \bibinfo{person}{Anthony Teachey}, {and} \bibinfo{person}{Brian Magerko}.} \bibinfo{year}{2022}\natexlab{}.
\newblock \showarticletitle{Family Learning Talk in AI Literacy Learning Activities}. In \bibinfo{booktitle}{\emph{Proceedings of the 2022 CHI Conference on Human Factors in Computing Systems}}. \bibinfo{pages}{1--20}.
\newblock


\bibitem[\protect\citeauthoryear{Lytle, Milliken, Catet{\'e}, and Barnes}{Lytle et~al\mbox{.}}{2020}]%
        {lytle2020investigating}
\bibfield{author}{\bibinfo{person}{Nicholas Lytle}, \bibinfo{person}{Alexandra Milliken}, \bibinfo{person}{Veronica Catet{\'e}}, {and} \bibinfo{person}{Tiffany Barnes}.} \bibinfo{year}{2020}\natexlab{}.
\newblock \showarticletitle{Investigating different assignment designs to promote collaboration in block-based environments}. In \bibinfo{booktitle}{\emph{Proceedings of the 51st ACM Technical Symposium on Computer Science Education}}. \bibinfo{pages}{832--838}.
\newblock


\bibitem[\protect\citeauthoryear{Margulieux, Dorn, and Searle}{Margulieux et~al\mbox{.}}{2019}]%
        {margulieux2019learning}
\bibfield{author}{\bibinfo{person}{Lauren~E Margulieux}, \bibinfo{person}{Brian Dorn}, {and} \bibinfo{person}{Kristin~A Searle}.} \bibinfo{year}{2019}\natexlab{}.
\newblock \showarticletitle{{Learning Sciences} for computing education}.
\newblock In \bibinfo{booktitle}{\emph{Cambridge Handbook of Computing Education Research}}, \bibfield{editor}{\bibinfo{person}{Sally~A Fincher} {and} \bibinfo{person}{Anthony~V Robins}} (Eds.). \bibinfo{publisher}{Cambridge: Cambridge University Press}.
\newblock


\bibitem[\protect\citeauthoryear{Marques, Gresse~von Wangenheim, and Hauck}{Marques et~al\mbox{.}}{2020}]%
        {marques2020teaching}
\bibfield{author}{\bibinfo{person}{L{\'\i}via~S Marques}, \bibinfo{person}{Christiane Gresse~von Wangenheim}, {and} \bibinfo{person}{Jean~CR Hauck}.} \bibinfo{year}{2020}\natexlab{}.
\newblock \showarticletitle{Teaching machine learning in school: A systematic mapping of the state of the art}.
\newblock \bibinfo{journal}{\emph{Informatics in Education}} \bibinfo{volume}{19}, \bibinfo{number}{2} (\bibinfo{year}{2020}), \bibinfo{pages}{283--321}.
\newblock


\bibitem[\protect\citeauthoryear{Mitchell, Baker, Moorosi, Denton, Hutchinson, Hanna, Gebru, and Morgenstern}{Mitchell et~al\mbox{.}}{2020}]%
        {mitchell2020diversity}
\bibfield{author}{\bibinfo{person}{Margaret Mitchell}, \bibinfo{person}{Dylan Baker}, \bibinfo{person}{Nyalleng Moorosi}, \bibinfo{person}{Emily Denton}, \bibinfo{person}{Ben Hutchinson}, \bibinfo{person}{Alex Hanna}, \bibinfo{person}{Timnit Gebru}, {and} \bibinfo{person}{Jamie Morgenstern}.} \bibinfo{year}{2020}\natexlab{}.
\newblock \showarticletitle{Diversity and inclusion metrics in subset selection}. In \bibinfo{booktitle}{\emph{Proceedings of the AAAI/ACM Conference on AI, Ethics, and Society}}. \bibinfo{pages}{117--123}.
\newblock


\bibitem[\protect\citeauthoryear{Noble}{Noble}{2018}]%
        {noble2018algorithms}
\bibfield{author}{\bibinfo{person}{Safiya~Umoja Noble}.} \bibinfo{year}{2018}\natexlab{}.
\newblock \bibinfo{booktitle}{\emph{Algorithms of oppression}}.
\newblock \bibinfo{publisher}{New York University Press}.
\newblock


\bibitem[\protect\citeauthoryear{of~Agriculture}{of~Agriculture}{2022}]%
        {Federal22}
\bibfield{author}{\bibinfo{person}{Department of Agriculture}.} \bibinfo{year}{2022}\natexlab{}.
\newblock \bibinfo{title}{Child Nutrition Programs: Income Eligibility Guidelines}.
\newblock
\newblock
\urldef\tempurl%
\url{https://www.govinfo.gov/content/pkg/FR-2022-02-16/pdf/2022-03261.pdf}
\showURL{%
\tempurl}


\bibitem[\protect\citeauthoryear{Payne}{Payne}{2019}]%
        {payne2019ethics}
\bibfield{author}{\bibinfo{person}{Blakeley~H Payne}.} \bibinfo{year}{2019}\natexlab{}.
\newblock \showarticletitle{An ethics of artificial intelligence curriculum for middle school students}.
\newblock  (\bibinfo{year}{2019}).
\newblock


\bibitem[\protect\citeauthoryear{Porter, Bouvier, Cutts, Grissom, Lee, McCartney, Zingaro, and Simon}{Porter et~al\mbox{.}}{2016}]%
        {porter2016multi}
\bibfield{author}{\bibinfo{person}{Leo Porter}, \bibinfo{person}{Dennis Bouvier}, \bibinfo{person}{Quintin Cutts}, \bibinfo{person}{Scott Grissom}, \bibinfo{person}{Cynthia Lee}, \bibinfo{person}{Robert McCartney}, \bibinfo{person}{Daniel Zingaro}, {and} \bibinfo{person}{Beth Simon}.} \bibinfo{year}{2016}\natexlab{}.
\newblock \showarticletitle{A multi-institutional study of peer instruction in introductory computing}. In \bibinfo{booktitle}{\emph{Proceedings of the 47th ACM Technical Symposium on Computing Science Education}}. \bibinfo{pages}{358--363}.
\newblock


\bibitem[\protect\citeauthoryear{Ramos, Meek, Simard, Suh, and Ghorashi}{Ramos et~al\mbox{.}}{2020}]%
        {ramos2020interactive}
\bibfield{author}{\bibinfo{person}{Gonzalo Ramos}, \bibinfo{person}{Christopher Meek}, \bibinfo{person}{Patrice Simard}, \bibinfo{person}{Jina Suh}, {and} \bibinfo{person}{Soroush Ghorashi}.} \bibinfo{year}{2020}\natexlab{}.
\newblock \showarticletitle{Interactive machine teaching: a human-centered approach to building machine-learned models}.
\newblock \bibinfo{journal}{\emph{Human--Computer Interaction}} \bibinfo{volume}{35}, \bibinfo{number}{5-6} (\bibinfo{year}{2020}), \bibinfo{pages}{413--451}.
\newblock


\bibitem[\protect\citeauthoryear{Redman}{Redman}{2018}]%
        {redman2018if}
\bibfield{author}{\bibinfo{person}{Thomas~C Redman}.} \bibinfo{year}{2018}\natexlab{}.
\newblock \showarticletitle{If your data is bad, your machine learning tools are useless}.
\newblock \bibinfo{journal}{\emph{Harvard Business Review}}  \bibinfo{volume}{2} (\bibinfo{year}{2018}).
\newblock


\bibitem[\protect\citeauthoryear{Register and Ko}{Register and Ko}{2020}]%
        {register2020learning}
\bibfield{author}{\bibinfo{person}{Yim Register} {and} \bibinfo{person}{Amy~J Ko}.} \bibinfo{year}{2020}\natexlab{}.
\newblock \showarticletitle{Learning machine learning with personal data helps stakeholders ground advocacy arguments in model mechanics}. In \bibinfo{booktitle}{\emph{Proceedings of the 2020 ACM Conference on International Computing Education Research}}. \bibinfo{pages}{67--78}.
\newblock


\bibitem[\protect\citeauthoryear{Rodr\'{\i}guez, Price, and Boyer}{Rodr\'{\i}guez et~al\mbox{.}}{2017}]%
        {exploringpairprogramming}
\bibfield{author}{\bibinfo{person}{Fernando~J. Rodr\'{\i}guez}, \bibinfo{person}{Kimberly~Michelle Price}, {and} \bibinfo{person}{Kristy~Elizabeth Boyer}.} \bibinfo{year}{2017}\natexlab{}.
\newblock \showarticletitle{Exploring the Pair Programming Process: Characteristics of Effective Collaboration}. In \bibinfo{booktitle}{\emph{Proceedings of the 2017 ACM SIGCSE Technical Symposium on Computer Science Education}} \emph{(\bibinfo{series}{SIGCSE '17})}. \bibinfo{publisher}{Association for Computing Machinery}, \bibinfo{address}{New York, NY, USA}, \bibinfo{pages}{507–512}.
\newblock
\showISBNx{9781450346986}
\urldef\tempurl%
\url{https://doi.org/10.1145/3017680.3017748}
\showDOI{\tempurl}


\bibitem[\protect\citeauthoryear{Roschelle}{Roschelle}{1992}]%
        {roschelle1992learning}
\bibfield{author}{\bibinfo{person}{Jeremy Roschelle}.} \bibinfo{year}{1992}\natexlab{}.
\newblock \showarticletitle{Learning by collaborating: Convergent conceptual change}.
\newblock \bibinfo{journal}{\emph{The journal of the learning sciences}} \bibinfo{volume}{2}, \bibinfo{number}{3} (\bibinfo{year}{1992}), \bibinfo{pages}{235--276}.
\newblock


\bibitem[\protect\citeauthoryear{Roschelle and Teasley}{Roschelle and Teasley}{1995}]%
        {roschelle1995construction}
\bibfield{author}{\bibinfo{person}{Jeremy Roschelle} {and} \bibinfo{person}{Stephanie~D Teasley}.} \bibinfo{year}{1995}\natexlab{}.
\newblock \showarticletitle{The construction of shared knowledge in collaborative problem solving}. In \bibinfo{booktitle}{\emph{Computer supported collaborative learning}}. Springer, \bibinfo{pages}{69--97}.
\newblock


\bibitem[\protect\citeauthoryear{Salleh, Mendes, and Grundy}{Salleh et~al\mbox{.}}{2011}]%
        {salleh2011effects}
\bibfield{author}{\bibinfo{person}{Norsaremah Salleh}, \bibinfo{person}{Emilia Mendes}, {and} \bibinfo{person}{John Grundy}.} \bibinfo{year}{2011}\natexlab{}.
\newblock \showarticletitle{The effects of openness to experience on pair programming in a higher education context}. In \bibinfo{booktitle}{\emph{2011 24th IEEE-CS Conference on Software Engineering Education and Training (CSEE\&T)}}. IEEE, \bibinfo{pages}{149--158}.
\newblock


\bibitem[\protect\citeauthoryear{Sambasivan, Kapania, Highfill, Akrong, Paritosh, and Aroyo}{Sambasivan et~al\mbox{.}}{2021}]%
        {sambasivan2021everyone}
\bibfield{author}{\bibinfo{person}{Nithya Sambasivan}, \bibinfo{person}{Shivani Kapania}, \bibinfo{person}{Hannah Highfill}, \bibinfo{person}{Diana Akrong}, \bibinfo{person}{Praveen Paritosh}, {and} \bibinfo{person}{Lora~M Aroyo}.} \bibinfo{year}{2021}\natexlab{}.
\newblock \showarticletitle{``Everyone wants to do the model work, not the data work'': Data Cascades in High-Stakes AI}. In \bibinfo{booktitle}{\emph{proceedings of the 2021 CHI Conference on Human Factors in Computing Systems}}. \bibinfo{pages}{1--15}.
\newblock


\bibitem[\protect\citeauthoryear{Sanusi, Oyelere, and Omidiora}{Sanusi et~al\mbox{.}}{2022}]%
        {sanusi2022exploring}
\bibfield{author}{\bibinfo{person}{Ismaila~Temitayo Sanusi}, \bibinfo{person}{Solomon~Sunday Oyelere}, {and} \bibinfo{person}{Joseph~Olamide Omidiora}.} \bibinfo{year}{2022}\natexlab{}.
\newblock \showarticletitle{Exploring teachers' preconceptions of teaching machine learning in high school: A preliminary insight from Africa}.
\newblock \bibinfo{journal}{\emph{Computers and Education Open}}  \bibinfo{volume}{3} (\bibinfo{year}{2022}), \bibinfo{pages}{100072}.
\newblock


\bibitem[\protect\citeauthoryear{Selwyn-Smith, Anslow, Homer, and Wallace}{Selwyn-Smith et~al\mbox{.}}{2019}]%
        {selwyn2019co}
\bibfield{author}{\bibinfo{person}{Ben Selwyn-Smith}, \bibinfo{person}{Craig Anslow}, \bibinfo{person}{Michael Homer}, {and} \bibinfo{person}{James~R Wallace}.} \bibinfo{year}{2019}\natexlab{}.
\newblock \showarticletitle{Co-located collaborative block-based programming}. In \bibinfo{booktitle}{\emph{2019 IEEE Symposium on Visual Languages and Human-Centric Computing (VL/HCC)}}. IEEE, \bibinfo{pages}{107--116}.
\newblock


\bibitem[\protect\citeauthoryear{Shapiro and Fiebrink}{Shapiro and Fiebrink}{2019}]%
        {shapiro2019introduction}
\bibfield{author}{\bibinfo{person}{R~Benjamin Shapiro} {and} \bibinfo{person}{Rebecca Fiebrink}.} \bibinfo{year}{2019}\natexlab{}.
\newblock \bibinfo{title}{Introduction to the special section: Launching an agenda for research on learning machine learning}.
\newblock , \bibinfo{numpages}{6}~pages.
\newblock


\bibitem[\protect\citeauthoryear{Simard, Amershi, Chickering, Pelton, Ghorashi, Meek, Ramos, Suh, Verwey, Wang, et~al\mbox{.}}{Simard et~al\mbox{.}}{2017}]%
        {simard2017machine}
\bibfield{author}{\bibinfo{person}{Patrice~Y Simard}, \bibinfo{person}{Saleema Amershi}, \bibinfo{person}{David~M Chickering}, \bibinfo{person}{Alicia~Edelman Pelton}, \bibinfo{person}{Soroush Ghorashi}, \bibinfo{person}{Christopher Meek}, \bibinfo{person}{Gonzalo Ramos}, \bibinfo{person}{Jina Suh}, \bibinfo{person}{Johan Verwey}, \bibinfo{person}{Mo Wang}, {et~al\mbox{.}}} \bibinfo{year}{2017}\natexlab{}.
\newblock \showarticletitle{Machine teaching: A new paradigm for building machine learning systems}.
\newblock \bibinfo{journal}{\emph{arXiv preprint arXiv:1707.06742}} (\bibinfo{year}{2017}).
\newblock


\bibitem[\protect\citeauthoryear{Strauss and Corbin}{Strauss and Corbin}{1990}]%
        {strauss1990basics}
\bibfield{author}{\bibinfo{person}{Anselm Strauss} {and} \bibinfo{person}{Juliet Corbin}.} \bibinfo{year}{1990}\natexlab{}.
\newblock \bibinfo{booktitle}{\emph{Basics of qualitative research}}.
\newblock \bibinfo{publisher}{Sage publications}.
\newblock


\bibitem[\protect\citeauthoryear{Tang}{Tang}{2019}]%
        {tang2019empowering}
\bibfield{author}{\bibinfo{person}{Danny Tang}.} \bibinfo{year}{2019}\natexlab{}.
\newblock \emph{\bibinfo{title}{Empowering Novices to Understand and Use Machine Learning With Personalized Image Classification Models, Intuitive Analysis Tools, and MIT App Inventor}}.
\newblock \bibinfo{thesistype}{Ph.D. Dissertation}. \bibinfo{school}{Massachusetts Institute of Technology}.
\newblock


\bibitem[\protect\citeauthoryear{Tayi and Ballou}{Tayi and Ballou}{1998}]%
        {tayi1998examining}
\bibfield{author}{\bibinfo{person}{Giri~Kumar Tayi} {and} \bibinfo{person}{Donald~P Ballou}.} \bibinfo{year}{1998}\natexlab{}.
\newblock \showarticletitle{Examining data quality}.
\newblock \bibinfo{journal}{\emph{Commun. ACM}} \bibinfo{volume}{41}, \bibinfo{number}{2} (\bibinfo{year}{1998}), \bibinfo{pages}{54--57}.
\newblock


\bibitem[\protect\citeauthoryear{Tedre, Denning, and Toivonen}{Tedre et~al\mbox{.}}{2021}]%
        {tedre2021ct}
\bibfield{author}{\bibinfo{person}{Matti Tedre}, \bibinfo{person}{Peter Denning}, {and} \bibinfo{person}{Tapani Toivonen}.} \bibinfo{year}{2021}\natexlab{}.
\newblock \showarticletitle{CT 2.0}. In \bibinfo{booktitle}{\emph{Proceedings of the 21st Koli Calling International Conference on Computing Education Research}}. \bibinfo{pages}{1--8}.
\newblock


\bibitem[\protect\citeauthoryear{Touretzky, Gardner-McCune, and Seehorn}{Touretzky et~al\mbox{.}}{2022}]%
        {touretzky2022machine}
\bibfield{author}{\bibinfo{person}{David Touretzky}, \bibinfo{person}{Christina Gardner-McCune}, {and} \bibinfo{person}{Deborah Seehorn}.} \bibinfo{year}{2022}\natexlab{}.
\newblock \showarticletitle{Machine Learning and the Five Big Ideas in AI}.
\newblock \bibinfo{journal}{\emph{International Journal of Artificial Intelligence in Education}} (\bibinfo{year}{2022}), \bibinfo{pages}{1--34}.
\newblock


\bibitem[\protect\citeauthoryear{Touretzky, Gardner-McCune, Martin, and Seehorn}{Touretzky et~al\mbox{.}}{2019}]%
        {ai4k12}
\bibfield{author}{\bibinfo{person}{David~S. Touretzky}, \bibinfo{person}{Christina Gardner-McCune}, \bibinfo{person}{Fred Martin}, {and} \bibinfo{person}{Deborah Seehorn}.} \bibinfo{year}{2019}\natexlab{}.
\newblock \bibinfo{title}{K-12 Guidelines for Artificial Intelligence: What Students Should Know}.
\newblock
\newblock
\urldef\tempurl%
\url{https://github.com/touretzkyds/ai4k12/raw/master/documents/ISTE_2019_Presentation_website_final.pdf/}
\showURL{%
\tempurl}


\bibitem[\protect\citeauthoryear{Tseng, Murai, Freed, Gelosi, Ta, and Kawahara}{Tseng et~al\mbox{.}}{2021}]%
        {tseng2021plushpal}
\bibfield{author}{\bibinfo{person}{Tiffany Tseng}, \bibinfo{person}{Yumiko Murai}, \bibinfo{person}{Natalie Freed}, \bibinfo{person}{Deanna Gelosi}, \bibinfo{person}{Tung~D Ta}, {and} \bibinfo{person}{Yoshihiro Kawahara}.} \bibinfo{year}{2021}\natexlab{}.
\newblock \showarticletitle{PlushPal: Storytelling with Interactive Plush Toys and Machine Learning}. In \bibinfo{booktitle}{\emph{Interaction Design and Children}}. \bibinfo{pages}{236--245}.
\newblock


\bibitem[\protect\citeauthoryear{Vachovsky, Wu, Chaturapruek, Russakovsky, Sommer, and Fei-Fei}{Vachovsky et~al\mbox{.}}{2016}]%
        {vachovsky2016toward}
\bibfield{author}{\bibinfo{person}{Marie~E Vachovsky}, \bibinfo{person}{Grace Wu}, \bibinfo{person}{Sorathan Chaturapruek}, \bibinfo{person}{Olga Russakovsky}, \bibinfo{person}{Richard Sommer}, {and} \bibinfo{person}{Li Fei-Fei}.} \bibinfo{year}{2016}\natexlab{}.
\newblock \showarticletitle{Toward more gender diversity in CS through an artificial intelligence summer program for high school girls}. In \bibinfo{booktitle}{\emph{Proceedings of the 47th ACM technical symposium on computing science education}}. \bibinfo{pages}{303--308}.
\newblock


\bibitem[\protect\citeauthoryear{Van~Brummelen, Heng, and Tabunshchyk}{Van~Brummelen et~al\mbox{.}}{2021}]%
        {van2021teaching}
\bibfield{author}{\bibinfo{person}{Jessica Van~Brummelen}, \bibinfo{person}{Tommy Heng}, {and} \bibinfo{person}{Viktoriya Tabunshchyk}.} \bibinfo{year}{2021}\natexlab{}.
\newblock \showarticletitle{Teaching tech to talk: K-12 conversational artificial intelligence literacy curriculum and development tools}. In \bibinfo{booktitle}{\emph{Proceedings of the AAAI Conference on Artificial Intelligence}}, Vol.~\bibinfo{volume}{35}. \bibinfo{pages}{15655--15663}.
\newblock


\bibitem[\protect\citeauthoryear{Vartiainen, Toivonen, Jormanainen, Kahila, Tedre, and Valtonen}{Vartiainen et~al\mbox{.}}{2020}]%
        {vartiainen2020machine}
\bibfield{author}{\bibinfo{person}{Henriikka Vartiainen}, \bibinfo{person}{Tapani Toivonen}, \bibinfo{person}{Ilkka Jormanainen}, \bibinfo{person}{Juho Kahila}, \bibinfo{person}{Matti Tedre}, {and} \bibinfo{person}{Teemu Valtonen}.} \bibinfo{year}{2020}\natexlab{}.
\newblock \showarticletitle{Machine learning for middle-schoolers: Children as designers of machine-learning apps}. In \bibinfo{booktitle}{\emph{2020 IEEE Frontiers in Education Conference (FIE)}}. IEEE, \bibinfo{pages}{1--9}.
\newblock


\bibitem[\protect\citeauthoryear{Wang, Mittal, Brooks, and Oney}{Wang et~al\mbox{.}}{2019}]%
        {wang2019data}
\bibfield{author}{\bibinfo{person}{April~Yi Wang}, \bibinfo{person}{Anant Mittal}, \bibinfo{person}{Christopher Brooks}, {and} \bibinfo{person}{Steve Oney}.} \bibinfo{year}{2019}\natexlab{}.
\newblock \showarticletitle{How data scientists use computational notebooks for real-time collaboration}.
\newblock \bibinfo{journal}{\emph{Proceedings of the ACM on Human-Computer Interaction}} \bibinfo{volume}{3}, \bibinfo{number}{CSCW} (\bibinfo{year}{2019}), \bibinfo{pages}{1--30}.
\newblock


\bibitem[\protect\citeauthoryear{Weiss, Khoshgoftaar, and Wang}{Weiss et~al\mbox{.}}{2016}]%
        {weiss2016survey}
\bibfield{author}{\bibinfo{person}{Karl Weiss}, \bibinfo{person}{Taghi~M Khoshgoftaar}, {and} \bibinfo{person}{DingDing Wang}.} \bibinfo{year}{2016}\natexlab{}.
\newblock \showarticletitle{A survey of transfer learning}.
\newblock \bibinfo{journal}{\emph{Journal of Big data}} \bibinfo{volume}{3}, \bibinfo{number}{1} (\bibinfo{year}{2016}), \bibinfo{pages}{1--40}.
\newblock


\bibitem[\protect\citeauthoryear{Werner, Hanks, and McDowell}{Werner et~al\mbox{.}}{2004}]%
        {werner2004pair}
\bibfield{author}{\bibinfo{person}{Linda~L Werner}, \bibinfo{person}{Brian Hanks}, {and} \bibinfo{person}{Charlie McDowell}.} \bibinfo{year}{2004}\natexlab{}.
\newblock \showarticletitle{Pair-programming helps female computer science students}.
\newblock \bibinfo{journal}{\emph{Journal on Educational Resources in Computing (JERIC)}} \bibinfo{volume}{4}, \bibinfo{number}{1} (\bibinfo{year}{2004}), \bibinfo{pages}{4--es}.
\newblock


\bibitem[\protect\citeauthoryear{Williams, Ali, Devasia, DiPaola, Hong, Kaputsos, Jordan, and Breazeal}{Williams et~al\mbox{.}}{2022}]%
        {williams2022ai}
\bibfield{author}{\bibinfo{person}{Randi Williams}, \bibinfo{person}{Safinah Ali}, \bibinfo{person}{Nisha Devasia}, \bibinfo{person}{Daniella DiPaola}, \bibinfo{person}{Jenna Hong}, \bibinfo{person}{Stephen~P Kaputsos}, \bibinfo{person}{Brian Jordan}, {and} \bibinfo{person}{Cynthia Breazeal}.} \bibinfo{year}{2022}\natexlab{}.
\newblock \showarticletitle{AI+ ethics curricula for middle school youth: Lessons learned from three project-based curricula}.
\newblock \bibinfo{journal}{\emph{International Journal of Artificial Intelligence in Education}} (\bibinfo{year}{2022}), \bibinfo{pages}{1--59}.
\newblock


\bibitem[\protect\citeauthoryear{Wilson and Daugherty}{Wilson and Daugherty}{[n. d.]}]%
        {smalldata2020}
\bibfield{author}{\bibinfo{person}{H.~James Wilson} {and} \bibinfo{person}{Paul~R. Daugherty}.} \bibinfo{year}{[n. d.]}\natexlab{}.
\newblock \bibinfo{title}{Small Data Can Play a Big Role in AI}.
\newblock
\newblock


\bibitem[\protect\citeauthoryear{Yang, Suh, Chen, and Ramos}{Yang et~al\mbox{.}}{2018}]%
        {yang2018grounding}
\bibfield{author}{\bibinfo{person}{Qian Yang}, \bibinfo{person}{Jina Suh}, \bibinfo{person}{Nan-Chen Chen}, {and} \bibinfo{person}{Gonzalo Ramos}.} \bibinfo{year}{2018}\natexlab{}.
\newblock \showarticletitle{Grounding interactive machine learning tool design in how non-experts actually build models}. In \bibinfo{booktitle}{\emph{Proceedings of the 2018 designing interactive systems conference}}. \bibinfo{pages}{573--584}.
\newblock


\bibitem[\protect\citeauthoryear{You}{You}{[n. d.]}]%
        {goodOnYou}
\bibfield{author}{\bibinfo{person}{Good~On You}.} \bibinfo{year}{[n. d.]}\natexlab{}.
\newblock
\newblock
\urldef\tempurl%
\url{https://goodonyou.eco/}
\showURL{%
\tempurl}


\bibitem[\protect\citeauthoryear{Zhou, Van~Brummelen, and Lin}{Zhou et~al\mbox{.}}{2020}]%
        {zhou2020designing}
\bibfield{author}{\bibinfo{person}{Xiaofei Zhou}, \bibinfo{person}{Jessica Van~Brummelen}, {and} \bibinfo{person}{Phoebe Lin}.} \bibinfo{year}{2020}\natexlab{}.
\newblock \showarticletitle{Designing AI Learning Experiences for K-12: Emerging Works, Future Opportunities and a Design Framework}.
\newblock \bibinfo{journal}{\emph{arXiv preprint arXiv:2009.10228}} (\bibinfo{year}{2020}).
\newblock


\bibitem[\protect\citeauthoryear{Zimmermann-Niefield, Polson, Moreno, and Shapiro}{Zimmermann-Niefield et~al\mbox{.}}{2020}]%
        {zimmermann2020youth}
\bibfield{author}{\bibinfo{person}{Abigail Zimmermann-Niefield}, \bibinfo{person}{Shawn Polson}, \bibinfo{person}{Celeste Moreno}, {and} \bibinfo{person}{R~Benjamin Shapiro}.} \bibinfo{year}{2020}\natexlab{}.
\newblock \showarticletitle{Youth making machine learning models for gesture-controlled interactive media}. In \bibinfo{booktitle}{\emph{Proceedings of the Interaction Design and Children Conference}}. \bibinfo{pages}{63--74}.
\newblock


\bibitem[\protect\citeauthoryear{Zimmermann-Niefield, Turner, Murphy, Kane, and Shapiro}{Zimmermann-Niefield et~al\mbox{.}}{2019}]%
        {zimmermann2019youth}
\bibfield{author}{\bibinfo{person}{Abigail Zimmermann-Niefield}, \bibinfo{person}{Makenna Turner}, \bibinfo{person}{Bridget Murphy}, \bibinfo{person}{Shaun~K Kane}, {and} \bibinfo{person}{R~Benjamin Shapiro}.} \bibinfo{year}{2019}\natexlab{}.
\newblock \showarticletitle{Youth learning machine learning through building models of athletic moves}. In \bibinfo{booktitle}{\emph{Proceedings of the 18th ACM International Conference on Interaction Design and Children}}. \bibinfo{pages}{121--132}.
\newblock


\end{thebibliography}

\appendix

\end{document}